\documentclass[aps,prx,groupedaddress,twocolumn,notitlepage,superscriptaddress,10pt]{revtex4-1}

\usepackage[T1]{fontenc}

\usepackage{float}
\usepackage{graphicx}  
\usepackage{xcolor}
\usepackage{bm}        
\usepackage{braket}
 \usepackage{amssymb}   
\usepackage{epstopdf}
\usepackage{xfrac}
\usepackage{cancel}
\usepackage{soul}
\usepackage{amsmath}
\usepackage{amsthm}
\usepackage{amsfonts}
\usepackage{bbm}
\definecolor{darkblue}{rgb}{0.1,0.2,0.6}
\definecolor{darkred}{rgb}{0.8,0.1,0.2}
\definecolor{darkgreen}{rgb}{0.31,0.62,0.24}
\usepackage[colorlinks,citecolor=darkblue,linkcolor=darkblue,urlcolor=darkblue]{hyperref} 
\usepackage{enumerate}
\usepackage{url}  
\usepackage{mathrsfs}
 \usepackage{mathtools}

\usepackage{multirow}
\usepackage{hhline}

\newcommand{\cz}[1]{\textsc{cz}_{#1}}

\usepackage[ruled, vlined, linesnumbered]{algorithm2e}
\newenvironment{algo}[1]{
  \algorithm[ht]
    \caption{#1}
    \DontPrintSemicolon
    \SetAlgoCaptionLayout{left}
    \SetAlgoHangIndent{0pt}
    \SetKwInOut{Input}{input}
    \SetKwInOut{Output}{output} 
}{
  \endalgorithm
}

\begin{document}

\title{All-photonic one-way quantum repeaters} 

\author{Daoheng Niu}
\affiliation{Cisco Quantum Lab, San Jose, CA 95134, USA}
\affiliation{Department of Physics, The University of Texas at Austin, Austin, TX 78712, USA}

\author{Yuxuan Zhang}
\affiliation{Cisco Quantum Lab, San Jose, CA 95134, USA}
\affiliation{Department of Physics, The University of Texas at Austin, Austin, TX 78712, USA}

\author{Alireza Shabani}
\affiliation{Cisco Quantum Lab, Los Angeles, CA 90049, USA}

\author{Hassan Shapourian}
\affiliation{Cisco Quantum Lab, San Jose, CA 95134, USA}

\begin{abstract}

Quantum repeater is the key technology enabler for long-distance quantum communication. To date, most of the existing quantum repeater protocols are designed based on specific quantum codes or graph states. In this paper, we propose a general framework for all-photonic one-way quantum repeaters based on the measurement-based error correction, which can be adapted to any Calderbank-Shor-Steane code including the recently discovered quantum low density parity check (QLDPC) codes. We present a novel decoding scheme, where the error correction process is carried out at the destination based on the accumulated data from the measurements made across the network. This procedure not only outperforms the conventional protocols with independent repeaters but also simplifies the local quantum operations at repeaters. As an example, we numerically show that the $[[48,6,8]]$ generalized bicycle code (as a small but efficient QLDPC code) has an equally good performance while reducing the resources  by at least an order of magnitude.

\end{abstract}

\maketitle

\section{Introduction}

Quantum network is one of the key quantum technologies and plays a central role in enabling unconditionally secure communication, distributed quantum computing, and quantum sensing~\cite{kimble2008quantum,wehner2018quantum}. Being an active area of research, the exact requirements and applications of a large-scale quantum network remain to be better understood. At the fundamental level nevertheless, a putative quantum network needs to provide a way for quantum communication, i.e., transfer of quantum information, among different network nodes where photons constitute the medium of choice.
Realizing a large-scale quantum network requires transmitting quantum information over long distances, that is challenging due to the photon loss which grows exponentially with distance. To circumvent this issue, quantum repeaters have been proposed~\cite{briegel1998quantum}, and there have been tremendous efforts over the past decade~\cite{munro2015inside,munro2012quantum,glaudell2016serialized,muralidharan2016optimal,pant2017rate,fowler2010surface,azuma2015all,lee2019fundamental,muralidharan2014ultrafast,namiki2016role,muralidharan2018oneway,ewert2016ultrafast,ewert2017ultrafast,borregaard2020one,zwerger2016measurement}.
The basic idea is to place a number of repeater {stations} at intermediate distances and use quantum correlations in multi-qubit entangled states to effectively enhance the transmission rate between two distant nodes.

Quantum repeater protocols are generally divided into two categories: The first category~\cite{briegel1998quantum,munro2015inside} is based on the heralded quantum entanglement distribution, where a pairwise entanglement between adjacent repeater nodes is established so that a long-range entanglement between the end nodes can be achieved via the entanglement swapping, i.e., performing Bell state measurement at each intermediate node. Quantum information is then transferred via the quantum teleportation.   
The success of a teleportation attempt relies on successfully establishing entanglement links between neighboring nodes and performing Bell measurements. Hence, a two-way classical channel is required to communicate the success of both processes to the adjacent nodes for every iteration. Two-way communication limits the performance of these protocols and may necessitate long-lived quantum memories at repeater stations, although the latter requirement in principle can be relaxed in all-photonic schemes~\cite{azuma2015all,lee2019fundamental}. 
The second category of repeater protocols~\cite{muralidharan2014ultrafast,namiki2016role,muralidharan2018oneway,ewert2016ultrafast,ewert2017ultrafast,borregaard2020one} involves sending encoded quantum information in the form of multi-qubit loss tolerant states which are received and (typically) error corrected at intermediate repeater stations. Such protocols only involve one-way communication and hence their performance is not impacted by the two-way communication requirement in the first category. Furthermore, the one-way protocols are far more efficient than the two-way protocols when it comes to network traffic in a large scale quantum network.

\begin{figure}
    \centering
    \includegraphics[scale=0.68]{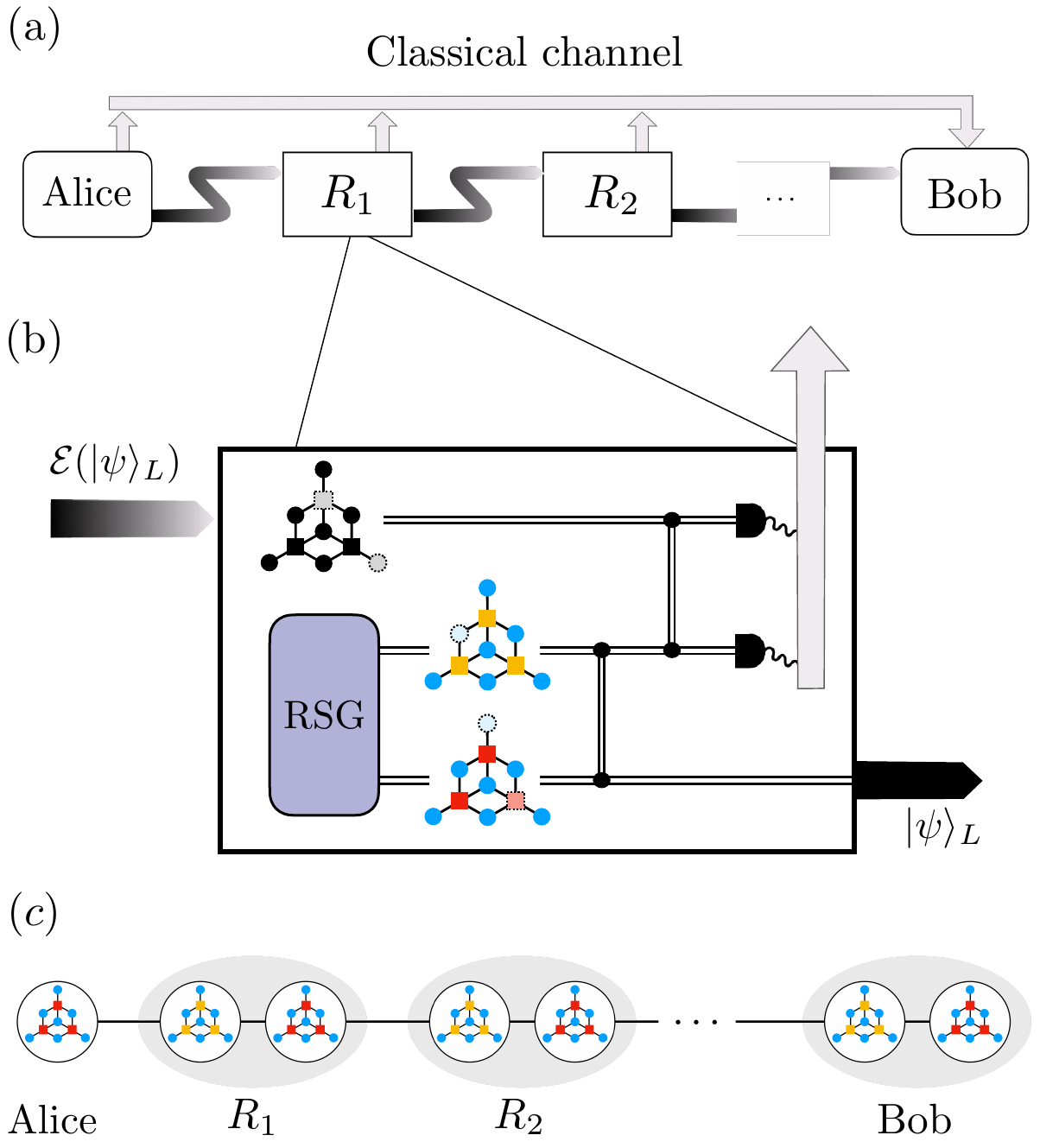}
    \caption{
    \textbf{All-photonic quantum repeater architecture} (a) Repeater chain, where $R_1$, $R_2$, $\cdots$ are the repeater nodes placed between Alice (sender) and Bob (receiver). Curved lines represent the quantum channel (optical fiber). (b) Inside a repeater station, where measurement-based error correction occurs. Two graph states associated with the $X$ (yellow) and $Z$ (red) stabilizers are generated at RSG with a transversal controlled-phase gate applied to them. The incoming logical qubit undergoes another controlled-phase gate with the graph state corresponding to the $X$ stabilizer. Measurement outcomes are relayed to the classical channel.
    Here, the $[[7,1,3]]$ Steane code is used for illustration purposes, where missing qubits are shown in light colors encircled with dashed lines (See Fig.~\ref{fig:foliated} for further information on graph state representation). (c) Syndrome graph for the error correction is realized at Bob's location and effectively forms a linear cluster state concatenated by the CSS stabilizer code (a.k.a.~foliated quantum code~\cite{bolt2016foliated}).}
    \label{fig:setup-schematics}
\end{figure}

In this paper, we introduce an all-photonic architecture for one-way quantum repeaters based on stabilizer codes realized by graph states of photons, where the photon loss is treated as a qubit erasure error and corrected through a measurement-based error correction scheme.
Our proposed architecture provides a general formalism that can be adapted to any Calderbank-Shor-Steane (CSS) stabilizer code.
In particular, one can leverage the remarkable properties (including large code distance) of the recently developed quantum low-density parity check (QLDPC) codes~\cite{panteleev2021degenerate,panteleev2021quantum} in this formalism.
We should contrast our repeater protocol with previous code-specific protocols such as those based on
the quantum parity code (QPC)~\cite{muralidharan2014ultrafast,namiki2016role,muralidharan2018oneway,ewert2016ultrafast,ewert2017ultrafast,lee2019fundamental},
where a teleportation-based error correction is performed to deal with erasure and possible operational errors, or other protocols based on tree graph states~\cite{varnava2006loss,zwerger2016measurement,borregaard2020one}, which can be viewed as teleportation path multiplexers.
Our repeater architecture in short involves encoding logical qubits in a graph state of photons corresponding to a CSS code and performing logical Bell state measurements at each repeater. The classical information obtained from measurement outcomes (which also contains loss events) is not processed until received by the recipient party who performs the error correction~\cite{raussendorf2006afault,bolt2016foliated} across the quantum network based on the accumulated data (See Fig.~\ref{fig:setup-schematics}). 
This feature is fundamentally different from conventional methods, where the error correction is performed at every repeater node,
and offers several advantages. First, 
the overall performance is improved over doing error correction at each repeater~\footnote{The intuitive reason behind this fact is that loss events can be corrected not only at the repeater node where the loss was observed but also through other stabilizers at neighboring repeater nodes.}.
Second, since there is no decoding at each repeater, there is no need for matter qubits and adaptive measurements. Third, for the same reason, the quantum gates and measurements within each repeater is independent of the choice of the stabilizer code. The latter two properties are in stark contrast with the previous studies where both the quantum and classical hardware as well as the error correction software were designed for specific encoding schemes such as the quantum parity code~\cite{muralidharan2014ultrafast,namiki2016role} or tree graph state~\cite{borregaard2020one}. This flexibility of our protocol would lead to a long-term advantage as the hardware technology is improved and new generation of quantum codes will be available. 

As we explain, the error correction in our architecture is effectively carried out on a one-dimensional cluster state concatenated by the CSS code as depicted in Fig.~\ref{fig:setup-schematics}(c). We derive the condition for a successful transmission of the logical states across the cluster state and provide a decoding algorithm.
We illustrate details of our framework using the $[[7,1,3]]$ Steane code~\cite{steane1996error} and $[[48,6,8]]$ generalized bicycle code~\cite{panteleev2021degenerate}, and numerically show that their performance is equal or better than existing protocols while requiring less resources.

\section{Results}

\subsection{Quantum repeater protocol}
\label{sec:protocol}

In this section, we introduce our quantum repeater architecture. As we explain, we use a measurement-based quantum error correction protocol so that the photon loss is treated as unheralded, and there is no need for long-lived matter-based quantum memory.
As shown in Fig.~\ref{fig:setup-schematics}, quantum information is encoded in a graph state realization of a quantum code, and repeaters are placed along the channel to correct errors occurring during the transmission through a lossy channel. Compared to existing quantum repeater proposals, our protocol does not require any extra quantum hardware overhead in addition to a resource-state generator (RSG) and single-photon detectors per each repeater. Furthermore, no classical data processing is required within the repeaters, and measurement outcomes are transmitted via a classical channel to the receiver.

We consider using single photons in the discrete-variable formalism such as time-bin encoding which is generally not sensitive to dephasing error and suitable for long-distance quantum communication~\cite{brendel1999pulsed}. 
The main source of error in our case is then photon loss 
which is detected during the measurement process and can be viewed as a quantum erasure channel where the error location is known but the error type is not. 
According to our protocol, the sender encodes the quantum information (logical qubits) in multi-photon graph states. There are two kinds of qubits in these graph states: data and ancilla qubits. Data qubits collectively encode the logical information, while ancilla qubits are used to measure the quantum code stabilizers. The size and shape of the graph are determined by the deployed CSS quantum code as will be explained in Sec.~\ref{sec:qecc}. 

At each repeater station (Fig.~\ref{fig:setup-schematics}(b)), upon receiving the incoming graph state (from the sender or previous repeater station), 
two graph states (associated with the $X$ and $Z$ stabilizer generators of the CSS code) are prepared by an RSG and form a (logical) Bell pair by applying a transversal controlled-phase gate. We note that this transversal gate can be incorporated into the state generation in the RSG.
Next, a transversal controlled-phase gate is applied between the received logical qubit and the local logical qubit corresponding to $X$ stabilizer; then, both qubits are sent to single-photon detectors where the physical qubits are all measured in $X$-basis. This step effectively teleports the input state to the remaining (unmeasured) local logical qubit which is transmitted to the next repeater. The teleportation process may seem to be reminiscent of the teleportation-based error correction schemes~\cite{knill2001scheme,knill2005quantum}; however,
the usage of CSS error correcting codes across the network results in a significantly greater loss tolerance. The key idea is that our decoder uses all the classical information obtained by measuring qubits across all repeaters as opposed to breaking it down to two qubit measurements per repeater. In other words, our measurement-based protocol leads to a loss tolerant channel by effectively realizing a linear cluster state of logical qubits between the sender and receiver as shown in Fig.~\ref{fig:setup-schematics}(c).

Our protocol corrects loss errors in the transmission via the optical fiber as well as during the state generation process.
In terms of their loss probability, there are two groups of photonic qubits: Those qubits which travel between the successive repeaters, and others which are generated and measured within a repeater. For instance, all ancilla qubits belong to the latter group.
The former qubits are subject to the erasure channel with an overall transmission probability,
\begin{align}
\label{eq: trans-prob}
    \eta (L) = \eta_r 10^{-\frac{\alpha_0}{10} L},
\end{align}
where $\alpha_0$ is the signal attenuation rate per unit length in the optical fiber (which we set to be $0.2$ dB/km and may report as $e^{-L/L_\text{att}}$ with the attenuation length of $L_{\text{att}} = 10/(\alpha_0 \ln(10))\approx 22$ km), 
$L$ is the travel distance, and  $\eta_r$ denotes the repeater efficiency  (or transmittance) which collectively includes
photon-source/detector efficiency, on-chip loss, and in/out coupling losses. Although the main contributing factor to the repeater efficiency depends on the details of generation/detection scheme, it is usually the case that 
in/out coupling at the chip-fiber interface is the dominant factor.
For this reason, we assume that the transmission efficiency of the latter group of qubits (i.e., internal qubits) is given by $\sqrt{\eta_r}$.

\begin{figure}
    \centering
    \includegraphics[scale=0.8]{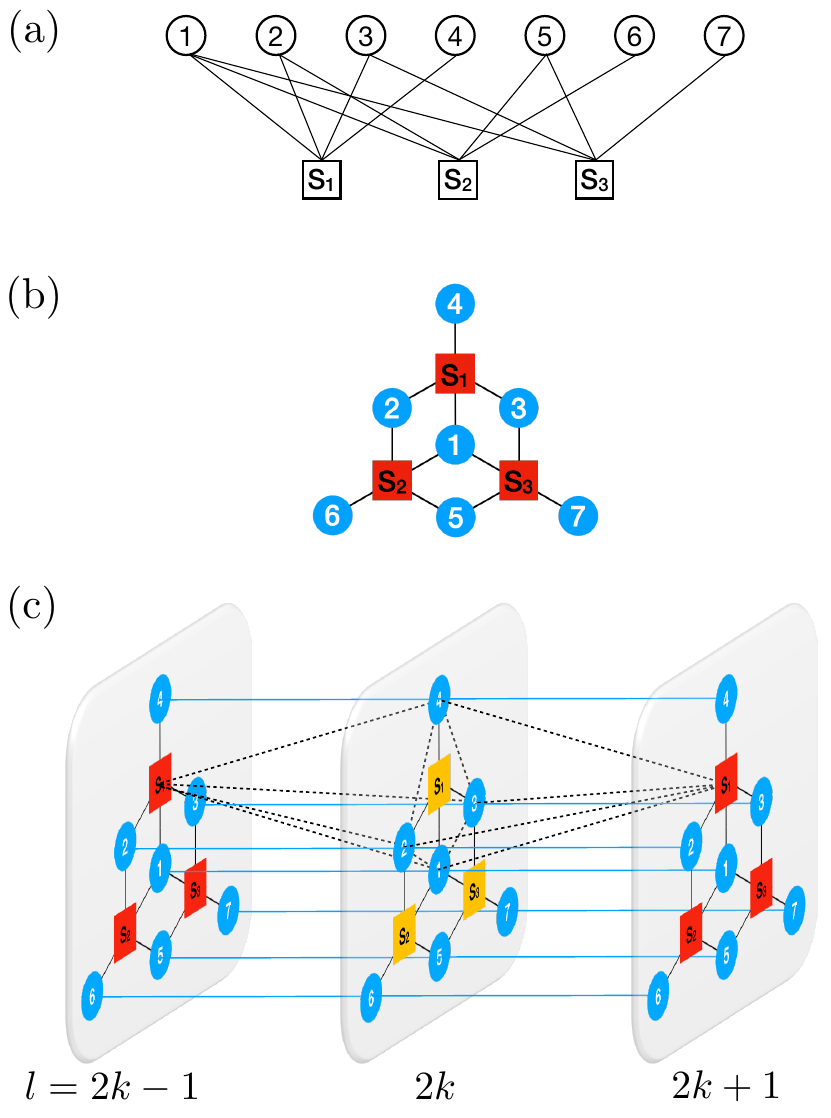}
    \caption{
    \textbf{Graph state construction of CSS codes} (a) Tanner graph of the parity-check matrix $H_X$ (or $H_Z$) and (b) the corresponding graph state for the $[[7, 1, 3]]$ Steane code. Circles with integer labels denote the data qubits and squares with labels $S_{1,2,3}$ denote the ancilla qubits to measure stabilizers. (c) Part of the graph state associated with the three consecutive sites on the 1D cluster state (Fig.~\ref{fig:setup-schematics}(c)). Ancilla qubits for two different sets of stabilizers are shown as red and yellow for $Z$ and $X$ stabilizers, respectively. In (b) and (c), we use graph state representation where solid lines represent controlled-phase gates. Dotted lines in (c) shows an example of an inter-site stabilizer operator.} 
    \label{fig:foliated} 
\end{figure}

\subsection{Measurement-based error correction}
\label{sec:qecc}

As mentioned, we use graph states to implement an all-photonic quantum code.
A graph state~(see Ref.~\cite{hein2006entanglement} for a detailed review) associated with graph $G$ of $N$ vertices (i.e., qubits) is defined as a quantum state of $N$ qubits 
 $   \ket{\Psi_G} =\prod_{(i,j)\in G} \cz{i,j} \ket{+}^{\otimes N},$
where subscripts are qubit labels $i,j=1,\cdots,N$, qubit state $\ket{+}$ denotes the eigenstate of $X$ Pauli operator, and $\cz{i,j}$ is a controlled-phase gate between qubits $i$ and $j$ for every edge $(i,j)$ on graph $G$. An important property of graph states is that they can be characterized as stabilizer states with $N$ stabilizer generators, where the $i$-th stabilizer generator associated with the $i$-th vertex on $G$ is defined by $P_i = X_i \bigotimes_{(i,j)\in G} Z_j$.
In other words, the $i$-th stabilizer is a product of the $X$ Pauli operator on $i$-th vertex and $Z$ Pauli operators on the adjacent (in the sense of graph) vertices. We should note that a graph state is a stabilizer state (as opposed to a stabilizer code) since there are $N$ stabilizers which determine a unique state for $N$ qubits.

A quantum code of distance $d$, denoted by $[[n, k, d]]$, encodes $k$ logical qubits into $n$ data qubits and is stabilized by $n-k$ Pauli operators (stabilizer generators or parity check operators). In the case of CSS codes, the stabilizer group is divided into two subgroups where the stabilizer operators are products of either only $X$ or $Z$ Pauli operators. The stabilizer group associated with $X$ or $Z$ operators can conveniently be represented by a bipartite graph (called Tanner graph) as shown for example in Fig.~\ref{fig:foliated}(a) for the $[[7,1,3]]$ Steane code. 
A straightforward implementation of a CSS quantum code in an all-photonic scheme is as follows: Construct the Tanner graph associated with $Z$ stabilizers as a graph state where parity check operators as well as data qubits are represented as vertices which we call ancilla and data qubits, respectively. By definition, measuring the ancilla qubits in $X$ basis then fixes the value of $Z$ parity checks (See e.g.~Fig.~\ref{fig:foliated}(b) for the 7-qubit code). Similarly, one can prepare a graph state (in Hadamard basis) in terms of the Tanner graph of $X$ operators and fix the $X$ parity checks by measuring the ancilla qubits.

As mentioned in the previous part, we prepare two graph states associated with $Z$ and $X$ check operators in each repeater and the receiver and apply controlled-phase gates to data qubits. This effectively realizes a linear cluster state of logical qubits (Fig.~\ref{fig:setup-schematics}(c)), where we run our error correction scheme to ensure that the input state from Alice is transmitted to Bob. In the remainder of this part, we explain the emergent stabilizer operators of the linear cluster state and discuss our decoding algorithm.
We denote the stabilizer groups and the logical operators of the underlying CSS quantum code by $\mathcal{S}_\sigma = \{ S_{\sigma,1},  S_{\sigma,2}, \cdots ,  S_{\sigma,(n-k)/2}\}$ and $\mathcal{L}_\sigma = \{ \tilde{\sigma}_1,\cdots,  \tilde{\sigma}_k \}$, respectively, where $\sigma=X,Z$ and tilde is added to distinguish the logical Pauli operators from the operators acting on the physical qubits.

Having $N-1$ repeater stations between Alice and Bob implies a $(2N+1)$-site cluster state where graph states in odd (even) layers are used for $Z$ ($X$) stabilizer measurements (Fig.~\ref{fig:setup-schematics}(c)). Each graph state consists of  $(3n-k)/2$ qubits where there are $n$ data qubits and $(n-k)/2$ ancilla qubits (Fig.~\ref{fig:foliated}(c)). We use superscripts to denote the site number and reserve subscripts to label data/ancilla qubits. For instance, $X_{q}^{(l)}$ with $q=1,\cdots, n$ and $X_{a,i}^{(l)}$ with $i = 1,\cdots,(n-k)/2$ refer to $X$ Pauli operators on data qubit $q$ and ancilla qubit $i$ on site $l$, respectively.
Following this notation, we write the on-site graph state stabilizers as 
\begin{align}
    G^{(l)}_{a,i} = 
    X_{a,i}^{(l)} \bigotimes_{q \in S_{\sigma^{(l)},i}} Z^{(l)}_q,
\end{align}
and inter-site graph state stabilizers as
\begin{align}
    G^{(l)}_{q}  = 
    X^{(l)}_q Z^{(l-1)}_q Z^{(l+1)}_q \bigotimes_{q \in S_{\sigma^{(l)},i}} Z_{a,i}^{(l)},
\end{align}
where 
$\sigma^{(2j)} = X$ and $\sigma^{(2j+1)} = Z$.
This gives overall $(2N+1)(3n-k)/2$ stabilizer generators.
We are measuring all physical qubits in $X$-basis, whereby the reduced stabilizer group can be constructed by combining data qubit stabilizers of a given site with the ancilla qubit stabilizers of its neighbors as shown by the dashed line in Fig.~\ref{fig:foliated}(c),
$P^{(l)}_i = G^{(l-1)}_{a,i} G^{(l+1)}_{a,i} \bigotimes_{q \in S_{\sigma^{(l+1)},i}} G^{(l)}_{q}$, 
which is further simplified into
\begin{align}
\label{eq:cluster-stabilizers}
    P^{(l)}_i = X^{(l-1)}_{a,i} X^{(l+1)}_{a,i} \bigotimes_{q \in S_{\bar\sigma^{(l)},i}} X^{(l)}_{q},
\end{align}
where we adopt the notation $\bar\sigma^{(l)} = H \sigma^{(l)} H$ to denote the Pauli operators after the Hadamard transformation (e.g., $\bar X = Z$ and $\bar Z = X$). 
The above simplification is due to the fact that $\mathcal{S}_X$ stabilizers commute with $\mathcal{S}_Z$ stabilizers; hence, ancilla qubits associated with $\bar \sigma$ stabilizers appear even number of times and cancel out.
We should note that there is no $X^{(l-1)}_{a,i}$ ($X^{(l+1)}_{a,i}$) in the above formula at the left (right) boundaries of the cluster state (c.f.~Fig.~\ref{fig:setup-schematics}(c)). There are $(2N+1)(n-k)$ stabilizer generators associated with $(n-k)$ ancilla qubits per site. Hence, the logical subspace contains $(2N+1)k$ logical qubits consistent with $(2N+1)$-site linear cluster state of $k$ logical qubits.

Our goal is to transfer logical qubits across the linear cluster state. To this end, the necessary logical operators to realize the measurement-based identity gate for  $i$-th logical qubit are given by
\begin{align}
    \mathcal{P}_{Z_i} &=\bigotimes_{l=2k}^{2N} \tilde X^{(l)}_i = \bigotimes_{l=2k}^{2N} \bigotimes_{q\in \tilde{X}_i} X_q^{(l)}, \nonumber \\
     \mathcal{P}_{X_i} &= \bigotimes_{l=2k-1}^{2N} \tilde X^{(l)}_i = \bigotimes_{l=2k-1}^{2N} \bigotimes_{q\in \tilde{X}_i} X_q^{(l)},
     \label{eq:cluster-logical}
\end{align}
which are the logical qubit version of the identity gate in measurement-based quantum computation~\cite{raussendorf2003measurement}.

As mentioned earlier, loss in our protocol is detected during the measurement process and is viewed as a quantum erasure channel. When it comes to applying controlled-phase gates, if a qubit is lost at an earlier point, then the gate will not be active.
Mathematically, loss error corresponds to partial tracing over the missing qubits. For instance, a loss (or erasure) channel for qubit $a$ is described by the following quantum channel,
\begin{align}
{\cal D}_a (\rho)  = \eta_a \rho + (1-\eta_a) \text{tr}_a(\rho) \otimes \ket{e}\!\bra{e},
\end{align}
where $\eta_a$ is the qubit transmission, $\ket{e}$ denotes an unknown state outside the computational basis for qubit $a$, which in our case correspond to an empty (vacuum) state with no photon in either bins. Alternatively, partial tracing is identical to conjugating with the Pauli group, because $\text{tr}_a(\rho)\otimes \openone_a = \frac{1}{4}(\rho+X_a\rho X_a+ Y_a\rho Y_a+Z_a\rho Z_a)$. Hence, after a loss event, only logical and stabilizer operators which commute with the Pauli operators of the erased qubits remain valid.
The loss tolerance is achieved in the following way: As long as the Pauli operators acting on the erased qubits  commute with the logical operators (\ref{eq:cluster-logical}) modulo the stabilizer group  (\ref{eq:cluster-stabilizers}), a successful transfer is guaranteed.

Lastly, we propose a simple decoding algorithm to determine the successful transmission of the input quantum states. 
First, we note that the stabilizer group (\ref{eq:cluster-stabilizers}) and the logical operators (\ref{eq:cluster-logical}) form two disjoint sets associated with 
even and odd sites (a.k.a., primal and dual syndrome graphs) which can be decoded independently. Then, for each set of stabilizers and logical operators the decoder checks whether or not logical operators can be combined with the stabilizer group such that they commute with Pauli operators acting on erased qubits.
The details are shown as pseudo codes in Algorithm~\ref{alg:decoder}.

\begin{algo}{Erasure decoder} 
\Input{A list of erased qubits $Q$.} 
\Output{Set of corrected qubits.}
\label{alg:decoder}
    \For{$q \in Q$}{
        \uIf{$|L| \neq 0$}{
            \For{$X_j \in L$}{
                \If{$X_j$ acts on $q$}{
                    Find $s_i \in S$ s.t.~$s_i$ acts on $q$.\;
                    {\bf If} {$s_i$ \emph{exists}} 
                         {\bf then} $X_j \gets s_i X_j$.\;
                    {\bf else} Remove $X_j$ from $L$.
                }
            }
            Find $R\subseteq S$ s.t.~$s_i\in R$ acts on $q$.\;
            {\bf If} {$|R|= 1$} {\bf then}
                Remove $s_i$ from $S$.\;
            \ElseIf{$|R|> 1$}{
                Construct stabilizer subgroup:  
                $\tilde R \gets \{s_i s_{i+1}|s_i\in R\}$.\;
                $S \gets (S\backslash R) \cup \tilde R $.}
        }
        {\bf else return} $\emptyset$.
    }
    {\bf return} $L$.
\end{algo}

\subsection{Performance and resource costs}
\label{sec:Performance and resource costs}

In this section, we investigate the performance of our proposed repeater architecture in terms of the effective transmission rate (ETR), denoted by $\eta_\text{eff}$, which is defined by the success probability of receiving the quantum information at the destination. 
Obviously, ETR has to outperform the direct transmission of single photons; however, placing too many repeaters at short distances from one another is expensive and not a scalable solution. 
Along this line, we discuss the trade-off between number of the repeaters and the ETR. We emphasize that our protocol can be developed for any CSS code. 
Intuitively, we look for $[[n,k,d]]$ codes with large code distance $d$, but we do not want to introduce too much overhead, i.e., we want the number of transmitted photons per logical qubits, $k/n$, to be small. Lifted product QLDPC codes~\cite{panteleev2021quantum,panteleev2021degenerate} with an almost linear distance $k,d\sim \Theta (n)$ could be a good candidate. However, we want practical solutions (at least for near-term implementations) where the size of on-site graph states in the 1D cluster state (c.f., Fig.~\ref{fig:setup-schematics}(c)), that is $(3n-k)/2$, is not too big. 

Given the above heuristic considerations, in what follows, we examine the performance of two example {CSS} codes: $[[7,1,3]]$ Steane code and $[[48,6,8]]$ generalized bicycle code. The first example represents a minimal quantum error correcting code and could be a good candidate (albeit with a limited performance) to consider for near-term experimental realization of our protocol. The second example is a representative quantum code with a decent code distance (which implies better performance) and moderate number of physical qubits. To investigate the performance of each implementation, we carry out Monte Carlo simulations where a random instance of erasure according to the loss probability for each type of qubits (data and ancilla) is generated and we use Algorithm~\ref{alg:decoder} to check if the given instance is correctable or not. We run this process over many iterations to accumulate statistics and evaluate the average success probability $\eta_\text{eff}$ (See Sec.~\ref{sec:mc} for further details).

\begin{figure}
    \centering
    \includegraphics[width=0.48\textwidth]{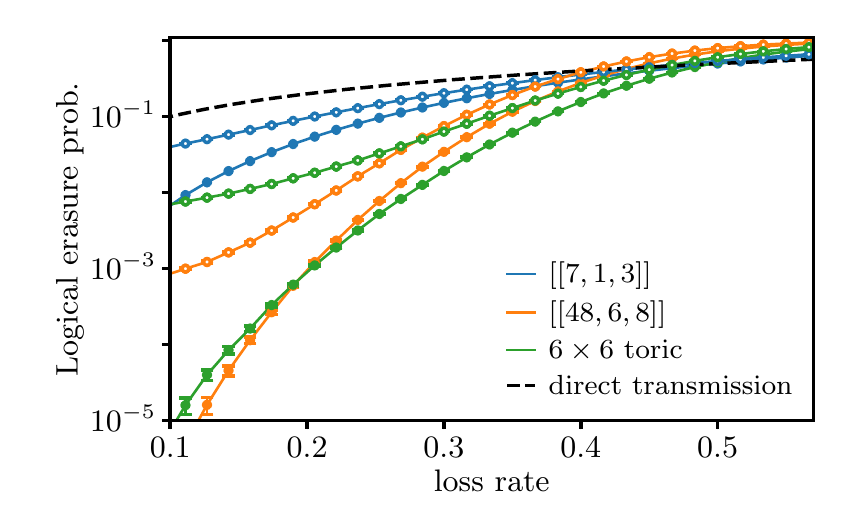}
    \caption{\textbf{Comparison of logical erasure probability of some representative CSS codes}  Loss tolerance of the $[[7,1,3]]$ Steane code, $[[48,6,8]]$ generalized bicycle code and $6\times 6$ toric code (as a reference) are plotted for two different repeater efficiencies $\eta_r = 1$ and $0.9$ which are shown as open and filled circles, respectively. The case of $\eta_r = 1$ for
    the $[[7,1,3]]$ code is analytically derived, as given in Eq.~(\ref{eq:7q-etr}). Each data point is averaged over $N_s = 2\times 10^6$ Monte Carlo iterations, and the error bars are given by $\sqrt{\eta_\text{eff}(1- \eta_\text{eff})/N_s}$.}
    \label{fig:no-repeater-performance}
\end{figure}

\begin{figure}
    \centering
    \includegraphics[width=0.45\textwidth]{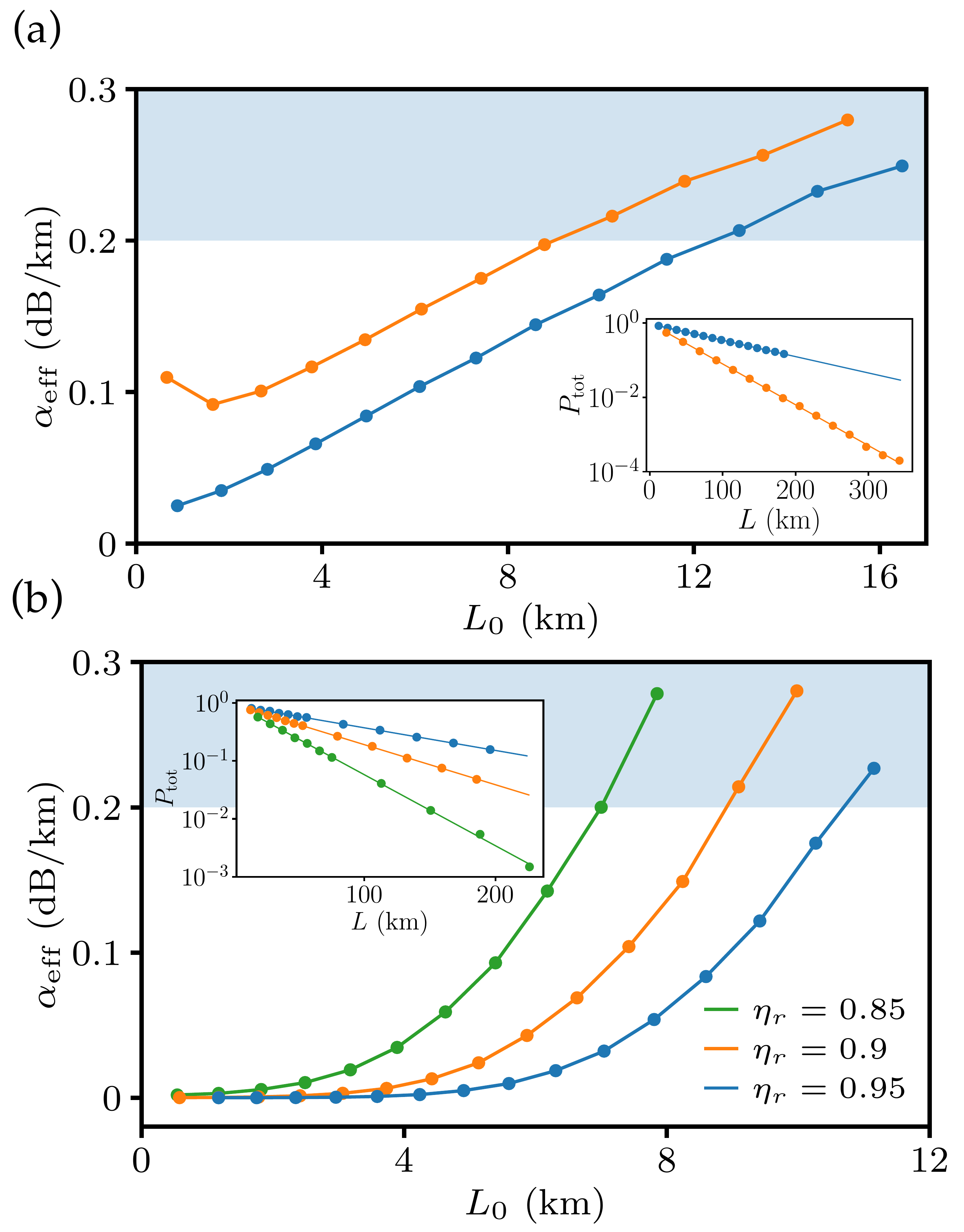}
    \caption{\textbf{Effective signal attenuation rate as a function of repeater spacing} Using Eq.~(\ref{eq:exp-fit}) the effective decay rate $\alpha_\text{eff}$ is obtained by fitting the data (see the insets for typical fits) and plotted against the distance between two neighboring quantum repeaters $L_0$ for (a) $[[7,1,3]]$ and (b) $[[48,6,8]]$ codes at different repeater efficiencies $\eta_r =$ $0.95$, $0.9$, and $0.85$. Error bars are not shown as they are comparable with the data marker size. The inset in (a) shows the effective transmission rate as a function of total distance $L$ for $(L_0,\eta_r) = (6.13\ \text{km}, 0.95)$  (blue) and $(11.41\ \text{km}, 0.9)$ (orange). The inset in (b) shows similar curves for $(7.0 \ \text{km},0.95)$ (blue), $(7.42 \ \text{km}, 0.9)$ (orange), and $(9.41\ \text{km}, 0.85)$ (green).
    Each data point is averaged over $N_s = 5\times 10^5$ Monte Carlo iterations.}
    \label{fig:decay-repeater-distance}
\end{figure}

We illustrate the loss tolerance of the two codes (without repeaters) in Fig.~\ref{fig:no-repeater-performance}, where for reference we also include the performance of the graph state implementation of $6\times 6$ toric code (i.e., $[[72,2,6]]$). 
This architecture corresponds to Fig.~\ref{fig:setup-schematics}(c) with only Alice and Bob and no repeaters, i.e., three-site cluster state.
We consider two cases: One is when there is no loss other than the channel attenuation $\eta_r=1$ (shown as filled circles), and  another case is when $\eta_r=0.9$ (shown as open circles). An immediate observation is that the break-even point for all encodings in the former case is nearly $0.5$; in other words, they outperform the direct transmission up to $50\%$ loss rate. The break-even point in the latter case is decreased as we consider repeater loss but they still perform better than the direct transition over a wide range of loss rates up to 0.4. We also observe that the performance of the $[[48,6,8]]$ code is close to or better than the toric code while the number of transmitted qubits per logical qubits is much less.

A simple way to characterize the overall performance of the repeater protocol is to calculate the effective signal attenuation $\alpha_\text{eff}$ as a function of the repeater spacing. To this end, we fit the ETR by an exponentially decaying function
\begin{align}
\label{eq:exp-fit}
    \eta_\text{eff}(L_0, L) \propto  10^{-\frac{\alpha_\text{eff}(L_0)}{10} L},
\end{align}
where  $L_0=L/N$ is the distance between two consecutive repeaters  out of total $(N-1)$ repeaters and plot them in Fig.~\ref{fig:decay-repeater-distance} in which the insets show some typical fits to the data. As expected, the performance of the repeater protocol, regardless of the underlying quantum code, degrades as we increase the repeater spacing.  It is worth noting that the effective attenuation of the $[[7,1,3]]$ code increases almost linearly with $L_0$ as opposed to that of the $[[48,6,8]]$ code. This can be attributed to the large code distance of the $[[48,6,8]]$ code compared to the $[[7,1,3]]$ code.
Furthermore, the shaded regions in the plots are for reference and indicate where the repeater scheme is no longer useful as it underperforms the direct transmission, i.e., $\alpha_\text{eff}>\alpha_0=0.2$ dB/km. We observe that the onset repeater spacing to enter the underperforming regime decreases with decreasing the repeater efficiency $\eta_r$. Another important observation is that as long as $L_0 \lesssim 4$~km the $[[48,6,8]]$ code  performs in an almost fully loss-tolerant regime where $\alpha_\text{eff}\approx 0$ despite the repeater photon loss of up to $10\%$.

\begin{figure*}
    \centering
    \includegraphics[width=0.9\textwidth]{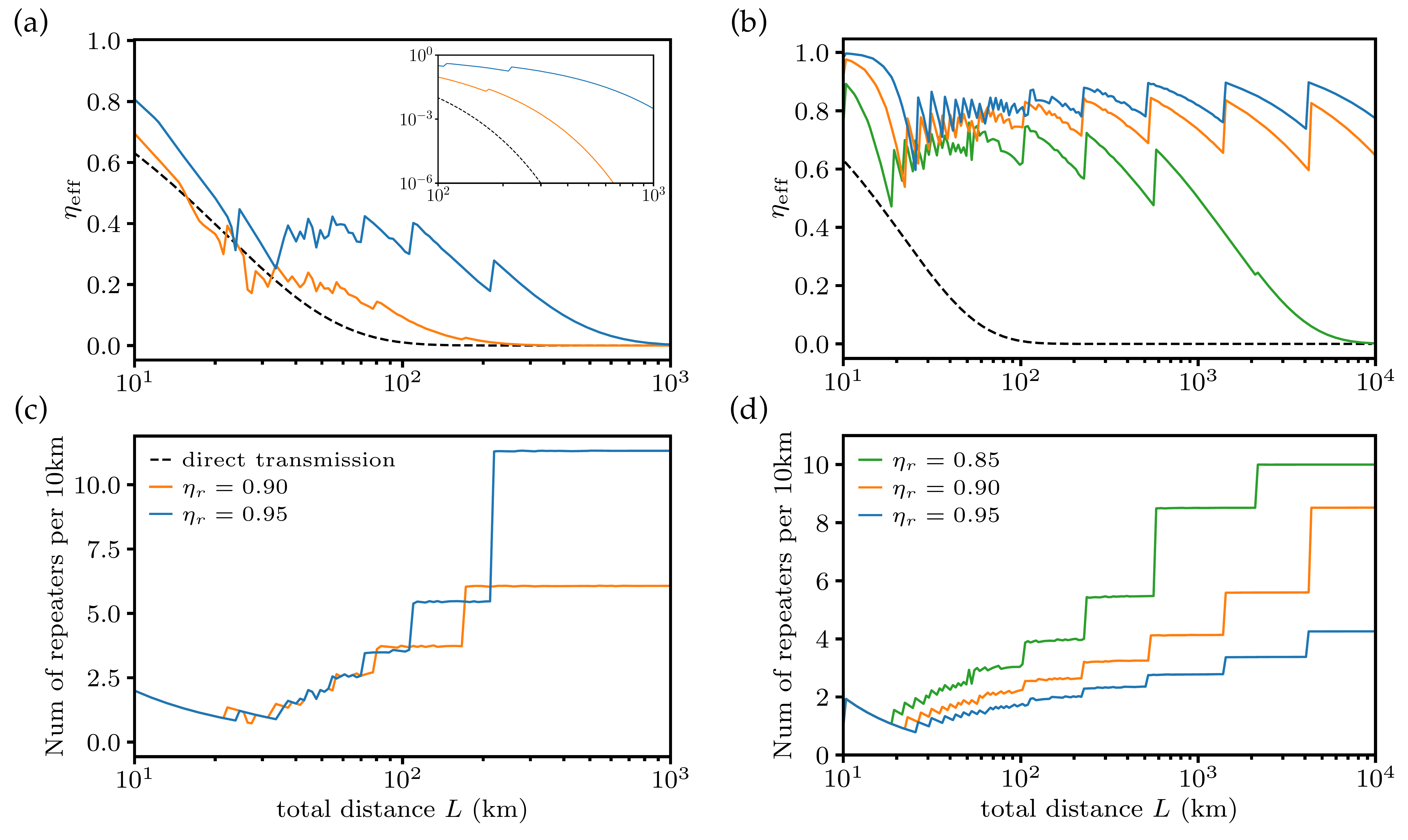}
    \caption{
    \textbf{Performance of optimized repeater architectures}
    (a) and (b) Effective transmission rate $\eta_{\text{eff}}$ of a repeater chain based on $[[7,1,3]]$ and $[[48,6,8]]$ codes after resource optimization (see main text for details) are plotted for different values of the repeater efficiency $\eta_r$. Inset of panel (a) shows the long distance behavior of $\eta_{\text{eff}}$ vs $L$ in logarithmic scale which manifestly outperforms the direct transmission above 200 km by at least three orders of magnitude. (c) and (d) Optimal number of repeaters per $10$ km as a function of total distance obtained by minimizing the cost function.     }
    \label{fig:optimized-arrangement-performance}
\end{figure*}

As mentioned before, repeater stations can be placed close to each other to obtain a total ETR of almost unity. However, this is not a practical approach for scalability and from the economic standpoint.
Therefore, we look for an optimization scheme to maximize the repeater spacing while not sacrificing the ETR as much. A natural choice for a cost function to be optimized is the ratio of the amount of resources used to the overall performance. More explicitly, we consider the ratio of the number of repeater stations per unit length to the total ETR. More details are explained in Sec.~\ref{sec:cost}. 
For a given repeater efficiency $\eta_r$ and total distance $L$, the ETR associated with the optimal repeater spacing is plotted  in Fig.~\ref{fig:optimized-arrangement-performance}. We note that the discontinuities in the optimal values is due to the discrete optimization over the number of repeaters.
As expected, the optimal number of repeaters per 10 km generally increases to compensate the decreasing ETR as the total distance increases. We also observe that the greater the repeater photon loss rate the closer the repeater nodes and the lower the ETR.

\begin{table*}
    \centering
{\footnotesize 
\renewcommand{\arraystretch}{1.2}
\begin{tabular}{lp{5cm}cccc}
    \hline
    Ref. & Characteristics & $\ \ \ \ \eta_r\ \ \ \ $ & ${k}/{n}$ & $\eta_\text{eff}$ & $L_0/L_\text{att}(\%)$ \\
    \hline
    \hline
    \multirow{2}{*}{Current work} & {Measurement-based error correction on CSS QEC graph states.} & \multirow{2}{*}{0.9} & \multirow{2}{*}{8} & \multirow{2}{*}{0.6-0.8} &  \multirow{2}{*}{4-14} \\ \hline
    \multirow{2}{*}{Muralidharan \emph{et al.~}\cite{muralidharan2014ultrafast}} & QPC based on teleportation-based error correction with matter qubits.  &  \multirow{2}{*}{0.9} &   \multirow{2}{*}{100-200} &     \multirow{2}{*}{0.6} &     \multirow{2}{*}{7.5} \\
    \hline
    \multirow{2}{*}{Ewert \emph{et al.~}\cite{ewert2016ultrafast}} & All-photonic teleportation error correction based on QPC.  &  \multirow{2}{*}{1} &   \multirow{2}{*}{100} &     \multirow{2}{*}{0.78} &     \multirow{2}{*}{10} \\
    \hline
    \multirow{1}{*}{Lee \emph{et al.~}\cite{lee2019fundamental}} & All-photonic QPC with linear optics.  &  \multirow{1}{*}{0.95} &   \multirow{1}{*}{400-500} &     \multirow{1}{*}{0.7} &     \multirow{1}{*}{4-9} \\
    \hline
    \multirow{2}{*}{Borregaard \emph{et al.~}\cite{borregaard2020one}} & {Tree graph state using quantum emitter and 2 auxilliary spin qubits} & \multirow{2}{*}{0.95} & \multirow{2}{*}{200-300} & \multirow{2}{*}{0.6} &     \multirow{2}{*}{13} \\ \hline
    \multirow{2}{*}{Rozp{\k{e}}dek \emph{et al.~}\cite{rozpedek2021quantum}} & Continuous-variable scheme based on GKP states requiring $18$dB squeezing.  &  \multirow{2}{*}{0.97} &   \multirow{2}{*}{4-7} &     \multirow{2}{*}{0.4-0.7} &     \multirow{2}{*}{1-1.3} \\
    \hline
\end{tabular}
}
\caption{Comparison with existing one-way protocols. The effective transmission rate $\eta_\text{eff}$ is reported for $10^3$-$10^4$ km. Here, $k/n$ denotes the number of photons per logical qubit, $\eta_r$ is the repeater efficiency, and $L_0/L_\text{att}$ is the ratio of the repeater spacing to the optical fiber attenuation length.}
    \label{tab:comparison}
\end{table*}

A few remarks are in order regarding the performance of our protocol compared to that of the existing protocols. As summarized in Table~\ref{tab:comparison}, the $[[48,6,8]]$ code delivers a similar value for ETR at $10^3$-$10^4$ km total communication distance while the repeater spacing (in units of the attenuation length $L_\text{att})$ is equal or greater. Furthermore, there is less constraint on minimum repeater efficiency $\eta_r$ and most importantly, an order of magnitude smaller number of photons per logical qubit $k/n$ than other discrete-variable protocols. We believe that the performance can even be improved further by utilizing more efficient QLDPC codes or larger codes with larger $n$ while keeping $k/n$ fixed.

\section{Discussion}

 In conclusion, we proposed
a general all-photonic one-way architecture for quantum repeaters which can be applied to any CSS code. We presented a novel error correction scheme based on measurement-based error correction, where one needs to only make projective measurements in a fixed basis at each repeater without further processing the outcomes, and the decoding process is performed at the destination. 
An immediate benefit of such error correction scheme is simplifying typical error correcting tasks involving feedforward processes in repeater nodes into some fixed operations such as controlled-phase gates and projective measurements in a fixed basis.
Moreover, error correction across the network has a better performance than the common approach where repeaters correct errors independently, because a local error can still be corrected along the repeater chain.
We study the performance of our architecture by using $[[7,1,3]]$ Steane code and $[[48,6,8]]$ generalized bicycle code. As we showed, the effective transmission rate at thousands of kilometers total distance is comparable with existing protocols at similar repeater spacings, while our protocol can work with an order of magnitude smaller number of photons per logical qubit and tolerate lower repeater efficiencies.

\begin{figure}
    \centering
    \includegraphics[scale=0.62]{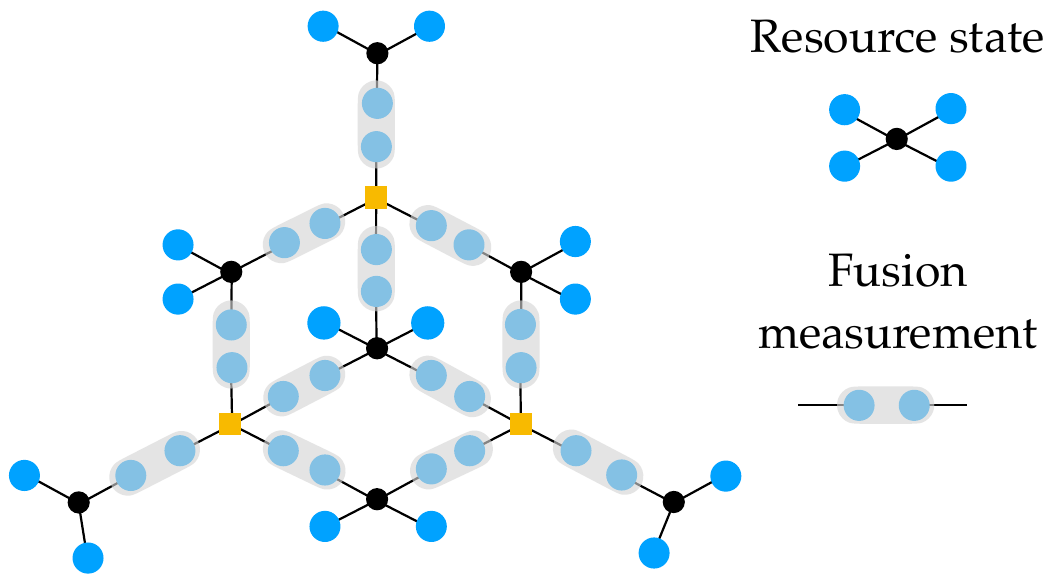}
    \caption{
    \textbf{Fusion-based realization of graph state} corresponding to the $[[7,1,3]]$ Steane code. The resource state is a star graph with 3 to 5 branches where the central vertex (shown as black circles or orange squares) could be a spin-qubit or photon which is measured during the generation process. There are two unmeasured data qubits per each vertex which are used to effectively implement the transversal controlled-phase gate between two adjacent graph states on the 1D cluster state.
    }
    \label{fig:fusion}
\end{figure}

In terms of hardware requirements at repeater stations, our architecture requires three components: a graph-state generator, a device for performing controlled-phase gate between the photonic qubits, and single-photon detectors, all of which are commonly used in previous architectures. Hence, we do not introduce any extra hardware overhead. 

As for the graph-state generator, various architectures based on solid-state quantum emitters coupled to nanophotonic waveguides have been proposed as deterministic generators~\cite{lindner2009proposal,buterakos2017deterministic,pichler2016photonic,pichler2017universal,wan2021fault,zhan2020deterministic,shapourian2022modular},
and there are ways to further optimize the graph-state generation algorithm~\cite{li2022photonic}.
Deterministic controlled-phase gates can possibly be realized by an array of (passive) quantum dots coupled to a waveguide~\cite{schrinski2021passive}.
Alternatively, the controlled-phase gate can be replaced by the fusion gates~\cite{browne2005resource,bartolucci2021fusion}, where small resource graph states are generated per each graph vertex and fusion gates are performed to construct a larger graph as shown in Fig.~\ref{fig:fusion}. Small resource states can be generated either deterministically based on quantum emitters or probabilistically using materials nonlinearity~\cite{wang2020integrated}. We note that the fusion gates based on linear optics are probabilistic and a fusion failure results in an erasure error (which ultimately reduces $\eta_r$). However, this constraint can be alleviated by increasing the success probability using boosted fusion gates~\cite{ewert2014efficient} with ancillary photons or inner encoding~\cite{bartolucci2021fusion},
or completely circumvented by using a different hardware, e.g., nearly-deterministic memory assisted Bell-state measurement devices~\cite{bhaskar2020experimental}.
In this regard, hybrid approaches to generate arbitrary graph states~\cite{hilaire2022near} can also be useful.
All in all, we envision a minimal hardware architecture based on few active elements where we use delay lines to generate time-multiplexed entangled photons.
We postpone a more detailed discussion on hardware designs and analysis of secret-key rate using a quantum key distribution scheme to a future work.

An important property of our architecture is being all-photonic where the major error is the qubit erasure which is handled more efficiently compared to the Pauli errors. This fact can be further leveraged by constructing quantum codes with a larger code distance and higher erasure threshold to design even more efficient and loss-tolerant protocols.
In a typical CSS code $(C_X,C_Z)$ corresponding to $X$ (primal graph) and $Z$ (dual graph) stabilizers, respectively, we want both $C_X$ and $C_Z$ codes to have a large code distance. However, our setup is asymmetric in the sense that only layers associated with primal graph ($C_X$) are subject to the channel loss. Therefore, it would be interesting to explore moderate-size CSS codes with a large code distance in one code (e.g., $C_X$) while the other code (e.g., $C_Z$) distance is not too large. 

Using the stabilizer formalism, we have derived the criterion for a successful transmission of logical qubits based on which we presented a simple quantum decoder. In principle, more efficient decoders ~\cite{delfosse2020linear,connolly2022fast} may not only improve the performance but also enable exploring many more codes and comparing their performance as a repeater protocol.

 Last but not least, the proposed quantum repeater architecture in this paper permits concatenation with other encoding schemes. For instance, one could replace
 each vertex in the graph state by a small tree graph to further enhance the performance. Moreover, similar architectures might be designed for continuous-variable encodings  where the graph vertices represent squeezed states or the Gottesman-Kitaev-Preskill (GKP) bosonic code,
 as it has been done for conventional repeater protocols~\cite{fukui2021all,rozpedek2021quantum,raveendran2022finite,Note5}.
 A unique advantage of the latter scheme is that deterministic fusion gates can be constructed in terms of beam splitters and homodyne measurements.
%
\footnotetext[5]{While such protocols feature no need for a quantum memory,  their realization is largely hindered by the difficulty of  generating GKP states in the optical domain with large squeezing values.}

\section{Methods}
\subsection{Monte Carlo Simulation}
\label{sec:mc}

As discussed in Sec.~\ref{sec:Performance and resource costs}, the performance of the quantum repeaters is determined by the transmission rate of the quantum information, which is defined as the average probability of the encoded logical qubit 
being successfully transmitted. 
We run Monte-Carlo simulations to evaluate the average probability given the number of repeaters. There is however some simplifications when the repeater efficiency is perfect, i.e., when $\eta_r = 1$; in this case, 
 the expression for the ETR is given by $\eta_{\text{eff}} = (\eta_{\text{eff},1})^N$ where $\eta_{\text{eff},1}$ is the ETR of directly transmitting the code.
This is because no loss for ancilla qubits implies measuring the stabilizers perfectly so each layer in Fig.~\ref{fig:foliated} operates independently. As a result, the overall ETR is given by the product of individual success probabilities.

It is possible to derive a closed-form expression for the success probability $\eta_{\text{eff},1}$ as a function of the photon direct transmission rate $\eta$ for small codes using basic combinatorics. For example, when considering the $[[7,1,3]]$ $\eta_{\text{eff},1}$ takes a polynomial form of order seven as follows
\begin{align}
\label{eq:7q-etr}
    \eta_{\text{eff},1} = \sum_{j=3}^7 a_j \eta^j (1-\eta)^{7-j},
\end{align}
where the coefficients are $(a_3,\cdots,a_7) = (7,28,21,7,1)$.
The above expression is plotted as the solid blue curve Fig.~\ref{fig:no-repeater-performance} which matches the numerical data points  shown as filled circles.

For more complex quantum codes or when we consider finite repeater efficiency $\eta_r < 1$, a closed-form expression cannot be easily obtained. Hence, we resort to numerical simulations. To this end, we perform Monte Carlo simulations and run our decoder on a $2N$-layer syndrome graph to determine how many logical qubits are successfully transmitted.
We note that the syndrome graph is decomposed into two $N$-layer subgraphs, a.k.a.~primal and dual, associated with $X$ and $Z$ stabilizers as expected for any CSS code (see  Sec.~\ref{sec:qecc} for more details).
Thus, each iteration involves psuedo-randomly generating a set of lost photons, where the loss events are independent and their probability depends on whether the qubit type is data or ancilla on primal or dual subgraph.
As explained in Sec.~\ref{sec:protocol}, the ancilla qubits on all layers as well as data qubits on even layers are not transmitted through the channel. As a result, we randomly generate loss events with probability $1-\sqrt{\eta_r}\approx (1-\eta_r)/2$ uniformly for all qubits on the dual syndrome subgraph (i.e., data qubits on even sites and ancilla qubits on odd sites) where the inter-site stabilizers are based on $H_Z$ parity-check matrix of the input CSS code. In contrast, we randomly generate loss events with probability $\eta$ in Eq.~(\ref{eq: trans-prob}) for data qubits on odd sites and with probability $1-\sqrt{\eta_r}$ for ancilla qubits on even sites to form the primal syndrome subgraph where the inter-site stabilizers are based on $H_X$ parity-check matrix.

Finally, the overall ETR $\eta_\text{eff}$ is found by multiplying the average success probability of the two syndrome graphs.
 This process constitutes how we generate the raw data; i.e., $\eta_\text{eff}$ as a function of number of repeaters, repeater efficiency $\eta_r$, and the channel loss rate. 
 To study the distance dependence of ETR and optimize the repeater spacing, we translate channel loss to distance via Eq.~(\ref{eq: trans-prob}) and interpolate or extrapolate the data using Eq.~(\ref{eq:exp-fit}) to obtain $\eta_\text{eff}$ for arbitrary distances.
 We justify this approximation numerically in Fig.~\ref{fig:decay-repeater-distance}.

\subsection{Cost function}
\label{sec:cost}

In this section, we discuss how to define a reasonable cost function which takes into account both the performance and the resource cost in our repeater architecture. For a fixed total distance $L$, as we increase the number of repeaters $N$, the repeater spacing $L_0$ is decreased, and the channel transmission rate $\eta$ approaches unity; as a result, the total ETR $\eta_\text{eff}$ saturates to its maximum value bounded by the repeater efficiency. However, this is not a cost-effective approach as quantum repeaters are costly resources. Therefore, we consider the trade-off between the performance and the resource cost. For this purpose, a quantitative measure of performance is simply the ETR. Moreover, the resource cost can be taken into account in terms of the number of repeaters per unit distance and the number of physical qubits in the deployed quantum code. This naturally leads to the following quantity for the cost function,
\begin{align}
    \label{eq:cost-func-1}
    C = \frac{N/ L}{\eta_{\text{eff}}} \cdot \frac{n_{\text{physical}}}{n_{\text{logical}}},
\end{align}
where $n_{\text{physical}}=n$, and $n_{\text{logical}}=k$ are the total number of physical qubits and the total number of logical qubits in a $[[n,k,d]]$ CSS code. We note that in our simulations we optimize the number of repeaters using the same code; hence, the second factor in $C$ is fixed in practice and the cost function only depends on the first factor. 

With the definition of cost function, we explore the optimized quantum repeaters arrangement over a wide range of total distances, from $L = 10$ km to $L = 10^{4}$ km. For a given $L$, we calculate the cost function over a range of numbers of repeaters from $1$ to $N_\text{max}$, where $N_\text{max}$ is chosen to be large enough so that the global minimum is contained. The resulting optimized number of repeaters $N_\text{opt}$  and the corresponding ETR as a function of total distance are plotted in Fig.~\ref{fig:optimized-arrangement-performance}. 
To evaluate $\eta_\text{eff}$ for arbitrary values of $N$ and $L_0$, we run Monte Carlo simulations on a 2D grid within the range $2\leq N \leq 30$ and $0 < \eta(L_0) < 1$, use the exponential ansatz (\ref{eq:exp-fit}) to extract $\alpha_\text{eff}(L_0)$ (along with the proportionality coefficient) as a function of $L_0$, and interpolate the corresponding value of effective decay rate. Finally, we plug in $N=L/L_0$ into Eq.~(\ref{eq:exp-fit}), obtain the ETR as well as the corresponding value for the cost function (\ref{eq:cost-func-1}). 

\section*{Data and Code Availability}

All the data presented in this paper is the result of numerical simulations. The code
used to generate this data is available upon
request.

\section*{Acknowledgements}

The authors acknowledge insightful discussions with Bing Qi, Galan Moody, Stefano Paesani, and Stephen DiAadamo. 

\section*{Competing interests}

The authors declare no competing interests.

\section*{Author Contribution}

HS conceived the project and designed the proposed repeater architectures. HS and DN designed and implemented the measurement-based error correction protocol and the Monte-Carlo simulations. YZ assisted with numerical implementation. HS and DN wrote the manuscript with inputs from all the coauthors.

\appendix
\section{QLPDC codes}
In this appendix, we briefly review the general properties of QLPDC codes and discuss the $[[48,6,8]]$ generalized bicycle code~\cite{panteleev2021degenerate} which is used as an example in our repeater protocol.

A stabilizer quantum code $[[n,k,d]]$ encodes $k$ logical qubits into $n$ ($n > k$) data qubits and is equipped with $(n-k)$ stabilizer generators, represented by the rows of the parity check binary matrix $H= (H_X | H_Z)$ where the columns are associated with $X$ and $Z$ Pauli operators acting on data qubits. A code is further characterized by the code distance $d$ which is related to its ability to correct errors. In our context, we use quantum codes to correct the qubit erasure error. For example, $d-1$ is the maximum number of erasure errors, which is guaranteed to be correctable. 
One specific type of quantum codes is the CSS codes, which we use in our quantum repeater architecture. Such codes are characterized by two classical linear codes $(C_X,C_Z)$ associated with $X$ and $Z$ stabilizers and their corresponding parity check matrices $H_X$ and $H_Z$. In other words, the parity-check matrix of a CSS code takes the following form, 
\begin{align}
H = \begin{bmatrix}
H_X &  \\
  & H_Z
\end{bmatrix}.
\end{align}
The commutation relation is equivalent to the binary identity  $H_X H_Z ^{T} = 0$. The number of logical qubits is found by $k = n - \text{rk}(H_X) - \text{rk}(H_Z)$, where one usually imposes the condition $\text{rk}(H_X) = \text{rk}(H_Z) = r$, such that $k=n-2r$. We use $\tilde{X}_i$ and $\tilde{Z}_i$ to denote the logical operator acting on the $i$-th logical qubit and represent them as rows of two binary matrices $L_X$ and $L_Z$, respectively. The fact that logical operators commute with the stabilizer generators while those acting on the same logical qubit anti-commute implies the following binary matrix identities \begin{align}
L_\sigma H_{\overline{\sigma}} = 0, \quad
L_\sigma L_{\sigma'}^T = \delta_{\overline{\sigma}\sigma'} \openone_k,
\end{align}
where $\sigma= X, Z$ denotes Pauli operators and $\overline \sigma= H\sigma H$ after the Hadamard transformation.
 Given a code with a parity-check matrix $H$, we use the recipe outlined in Refs.~\cite{gottesman1997stabilizer,Nielsen-Chuang} to find the logical operators based on an ansatz satisfying the above commutation relations. In short, we obtain the logical operators by first transforming $H$ into the standard form via Gaussian elimination and then solving for the ansatz.

A quantum low density parity check (QLDPC) code is a stabilizer code defined by a sparse binary parity-check matrix, where the sparsity implies that the weights of all rows and columns in $H$ are upper bounded by some universal constant as the code length $n$ grows in an infinite family of codes.
We use the $[[48,6,8]]$ code in our repeater protocol, which belongs to a family of codes called the generalized bicycle (GB) codes. The GB codes provide a general ansatz for CSS codes~\cite{Kovalev2013QLDPCfiniterate}, where the parity-check matrices are defined by $H_X = [A, B]$ and $H_Z = [B^{\text{T}}, A^{\text{T}}]$ in terms of two binary $\ell \times \ell$ circulant matrices $A$ and $B$ which always commute. We note that the commutation relation $H_X H^{T}_Z = 0$ is manifestly satisfied since $A$ and $B$ commute. A binary circulant matrix $A$ is represented as $A = a_0 I + a_1 P + ... + a_{\ell-1} P^{\ell-1}$ 
where $I$ is the $\ell\times \ell$ identity matrix and $P$ is the permutation matrix, i.e., the right cyclic shift by one position $P = [P_{ij}]_{\ell\times \ell}$ with $P_{ij}=\delta_{i-1,j}$ where $\delta_{ij}$ is the Kronecker delta function. 
A circulant matrix can alternatively be represented in a polynomial form $a(x) = a_0 + a_1 x + ... + a_{l-1} x^{l-1}$. 
An important property of circulant matrices is that their rank is determined algebraically as $\ell - \deg g(x)$ where $\deg g(x)$ is the degree of the polynomial $g(x) = \gcd(a(x),x^{\ell}-1)$. Therefore, the dimension of a GB code defined by $a(x)$ and $b(x)$ is given by $k =2 \deg g(x)$ where $g(x) = \gcd (a(x),b(x),x^\ell-1)$ because $\text{rk}(H_X) = \text{rk}(H_Z) = n- \deg g(x)$. 
As explained in Ref.~\cite{panteleev2021degenerate}, possible values of the dimension $k$ of the GB codes with a fixed circulant size $\ell$ correspond to the degree of $g(x)$, that is given by all possible factors of the polynomial $x^\ell - 1$. Hence, to produce a QLDPC code, one needs to find low-weight polynomials $a(x)$ and $b(x)$. This can be done via an exhaustive search over all polynomials of the given weight when $\ell$ is relatively small or done via a random search where a high success probability is guaranteed.

The $[[48,6,8]]$ code is characterized by $\ell=24$ and the following circulant matrices~\cite{panteleev2021quantum}: $a(x) = 1+ x^2 + x^8 + x^{15}$, $b(x) = 1 + x^2 + x^{12} + x^{17}$.

\bibliography{references.bib}

\begin{thebibliography}{54}%
\makeatletter
\providecommand \@ifxundefined [1]{%
 \@ifx{#1\undefined}
}%
\providecommand \@ifnum [1]{%
 \ifnum #1\expandafter \@firstoftwo
 \else \expandafter \@secondoftwo
 \fi
}%
\providecommand \@ifx [1]{%
 \ifx #1\expandafter \@firstoftwo
 \else \expandafter \@secondoftwo
 \fi
}%
\providecommand \natexlab [1]{#1}%
\providecommand \enquote  [1]{``#1''}%
\providecommand \bibnamefont  [1]{#1}%
\providecommand \bibfnamefont [1]{#1}%
\providecommand \citenamefont [1]{#1}%
\providecommand \href@noop [0]{\@secondoftwo}%
\providecommand \href [0]{\begingroup \@sanitize@url \@href}%
\providecommand \@href[1]{\@@startlink{#1}\@@href}%
\providecommand \@@href[1]{\endgroup#1\@@endlink}%
\providecommand \@sanitize@url [0]{\catcode `\\12\catcode `\$12\catcode
  `\&12\catcode `\#12\catcode `\^12\catcode `\_12\catcode `\%12\relax}%
\providecommand \@@startlink[1]{}%
\providecommand \@@endlink[0]{}%
\providecommand \url  [0]{\begingroup\@sanitize@url \@url }%
\providecommand \@url [1]{\endgroup\@href {#1}{\urlprefix }}%
\providecommand \urlprefix  [0]{URL }%
\providecommand \Eprint [0]{\href }%
\providecommand \doibase [0]{http://dx.doi.org/}%
\providecommand \selectlanguage [0]{\@gobble}%
\providecommand \bibinfo  [0]{\@secondoftwo}%
\providecommand \bibfield  [0]{\@secondoftwo}%
\providecommand \translation [1]{[#1]}%
\providecommand \BibitemOpen [0]{}%
\providecommand \bibitemStop [0]{}%
\providecommand \bibitemNoStop [0]{.\EOS\space}%
\providecommand \EOS [0]{\spacefactor3000\relax}%
\providecommand \BibitemShut  [1]{\csname bibitem#1\endcsname}%
\let\auto@bib@innerbib\@empty
\bibitem [{\citenamefont {Kimble}(2008)}]{kimble2008quantum}%
  \BibitemOpen
  \bibfield  {author} {\bibinfo {author} {\bibfnamefont {H.~J.}\ \bibnamefont
  {Kimble}},\ }\href {\doibase 10.1038/nature07127} {\bibfield  {journal}
  {\bibinfo  {journal} {Nature}\ }\textbf {\bibinfo {volume} {453}},\ \bibinfo
  {pages} {1023} (\bibinfo {year} {2008})}\BibitemShut {NoStop}%
\bibitem [{\citenamefont {Wehner}\ \emph {et~al.}(2018)\citenamefont {Wehner},
  \citenamefont {Elkouss},\ and\ \citenamefont {Hanson}}]{wehner2018quantum}%
  \BibitemOpen
  \bibfield  {author} {\bibinfo {author} {\bibfnamefont {S.}~\bibnamefont
  {Wehner}}, \bibinfo {author} {\bibfnamefont {D.}~\bibnamefont {Elkouss}}, \
  and\ \bibinfo {author} {\bibfnamefont {R.}~\bibnamefont {Hanson}},\ }\href
  {\doibase 10.1126/science.aam9288} {\bibfield  {journal} {\bibinfo  {journal}
  {Science}\ }\textbf {\bibinfo {volume} {362}},\ \bibinfo {pages} {eaam9288}
  (\bibinfo {year} {2018})}\BibitemShut {NoStop}%
\bibitem [{\citenamefont {Briegel}\ \emph {et~al.}(1998)\citenamefont
  {Briegel}, \citenamefont {D\"ur}, \citenamefont {Cirac},\ and\ \citenamefont
  {Zoller}}]{briegel1998quantum}%
  \BibitemOpen
  \bibfield  {author} {\bibinfo {author} {\bibfnamefont {H.-J.}\ \bibnamefont
  {Briegel}}, \bibinfo {author} {\bibfnamefont {W.}~\bibnamefont {D\"ur}},
  \bibinfo {author} {\bibfnamefont {J.~I.}\ \bibnamefont {Cirac}}, \ and\
  \bibinfo {author} {\bibfnamefont {P.}~\bibnamefont {Zoller}},\ }\href
  {\doibase 10.1103/PhysRevLett.81.5932} {\bibfield  {journal} {\bibinfo
  {journal} {Phys. Rev. Lett.}\ }\textbf {\bibinfo {volume} {81}},\ \bibinfo
  {pages} {5932} (\bibinfo {year} {1998})}\BibitemShut {NoStop}%
\bibitem [{\citenamefont {Munro}\ \emph {et~al.}(2015)\citenamefont {Munro},
  \citenamefont {Azuma}, \citenamefont {Tamaki},\ and\ \citenamefont
  {Nemoto}}]{munro2015inside}%
  \BibitemOpen
  \bibfield  {author} {\bibinfo {author} {\bibfnamefont {W.~J.}\ \bibnamefont
  {Munro}}, \bibinfo {author} {\bibfnamefont {K.}~\bibnamefont {Azuma}},
  \bibinfo {author} {\bibfnamefont {K.}~\bibnamefont {Tamaki}}, \ and\ \bibinfo
  {author} {\bibfnamefont {K.}~\bibnamefont {Nemoto}},\ }\href {\doibase
  10.1109/JSTQE.2015.2392076} {\bibfield  {journal} {\bibinfo  {journal} {IEEE
  Journal of Selected Topics in Quantum Electronics}\ }\textbf {\bibinfo
  {volume} {21}},\ \bibinfo {pages} {78} (\bibinfo {year} {2015})}\BibitemShut
  {NoStop}%
\bibitem [{\citenamefont {Munro}\ \emph {et~al.}(2012)\citenamefont {Munro},
  \citenamefont {Stephens}, \citenamefont {Devitt}, \citenamefont {Harrison},\
  and\ \citenamefont {Nemoto}}]{munro2012quantum}%
  \BibitemOpen
  \bibfield  {author} {\bibinfo {author} {\bibfnamefont {W.~J.}\ \bibnamefont
  {Munro}}, \bibinfo {author} {\bibfnamefont {A.~M.}\ \bibnamefont {Stephens}},
  \bibinfo {author} {\bibfnamefont {S.~J.}\ \bibnamefont {Devitt}}, \bibinfo
  {author} {\bibfnamefont {K.~A.}\ \bibnamefont {Harrison}}, \ and\ \bibinfo
  {author} {\bibfnamefont {K.}~\bibnamefont {Nemoto}},\ }\href {\doibase
  10.1038/nphoton.2012.243} {\bibfield  {journal} {\bibinfo  {journal} {Nature
  Photonics}\ }\textbf {\bibinfo {volume} {6}},\ \bibinfo {pages} {777}
  (\bibinfo {year} {2012})}\BibitemShut {NoStop}%
\bibitem [{\citenamefont {Glaudell}\ \emph {et~al.}(2016)\citenamefont
  {Glaudell}, \citenamefont {Waks},\ and\ \citenamefont
  {Taylor}}]{glaudell2016serialized}%
  \BibitemOpen
  \bibfield  {author} {\bibinfo {author} {\bibfnamefont {A.~N.}\ \bibnamefont
  {Glaudell}}, \bibinfo {author} {\bibfnamefont {E.}~\bibnamefont {Waks}}, \
  and\ \bibinfo {author} {\bibfnamefont {J.~M.}\ \bibnamefont {Taylor}},\
  }\href {\doibase 10.1088/1367-2630/18/9/093008} {\bibfield  {journal}
  {\bibinfo  {journal} {New Journal of Physics}\ }\textbf {\bibinfo {volume}
  {18}},\ \bibinfo {pages} {093008} (\bibinfo {year} {2016})}\BibitemShut
  {NoStop}%
\bibitem [{\citenamefont {Muralidharan}\ \emph {et~al.}(2016)\citenamefont
  {Muralidharan}, \citenamefont {Li}, \citenamefont {Kim}, \citenamefont
  {L{\"u}tkenhaus}, \citenamefont {Lukin},\ and\ \citenamefont
  {Jiang}}]{muralidharan2016optimal}%
  \BibitemOpen
  \bibfield  {author} {\bibinfo {author} {\bibfnamefont {S.}~\bibnamefont
  {Muralidharan}}, \bibinfo {author} {\bibfnamefont {L.}~\bibnamefont {Li}},
  \bibinfo {author} {\bibfnamefont {J.}~\bibnamefont {Kim}}, \bibinfo {author}
  {\bibfnamefont {N.}~\bibnamefont {L{\"u}tkenhaus}}, \bibinfo {author}
  {\bibfnamefont {M.~D.}\ \bibnamefont {Lukin}}, \ and\ \bibinfo {author}
  {\bibfnamefont {L.}~\bibnamefont {Jiang}},\ }\href {\doibase
  10.1038/srep20463} {\bibfield  {journal} {\bibinfo  {journal} {Scientific
  reports}\ }\textbf {\bibinfo {volume} {6}},\ \bibinfo {pages} {1} (\bibinfo
  {year} {2016})}\BibitemShut {NoStop}%
\bibitem [{\citenamefont {Pant}\ \emph {et~al.}(2017)\citenamefont {Pant},
  \citenamefont {Krovi}, \citenamefont {Englund},\ and\ \citenamefont
  {Guha}}]{pant2017rate}%
  \BibitemOpen
  \bibfield  {author} {\bibinfo {author} {\bibfnamefont {M.}~\bibnamefont
  {Pant}}, \bibinfo {author} {\bibfnamefont {H.}~\bibnamefont {Krovi}},
  \bibinfo {author} {\bibfnamefont {D.}~\bibnamefont {Englund}}, \ and\
  \bibinfo {author} {\bibfnamefont {S.}~\bibnamefont {Guha}},\ }\href {\doibase
  10.1103/PhysRevA.95.012304} {\bibfield  {journal} {\bibinfo  {journal} {Phys.
  Rev. A}\ }\textbf {\bibinfo {volume} {95}},\ \bibinfo {pages} {012304}
  (\bibinfo {year} {2017})}\BibitemShut {NoStop}%
\bibitem [{\citenamefont {Fowler}\ \emph {et~al.}(2010)\citenamefont {Fowler},
  \citenamefont {Wang}, \citenamefont {Hill}, \citenamefont {Ladd},
  \citenamefont {Van~Meter},\ and\ \citenamefont
  {Hollenberg}}]{fowler2010surface}%
  \BibitemOpen
  \bibfield  {author} {\bibinfo {author} {\bibfnamefont {A.~G.}\ \bibnamefont
  {Fowler}}, \bibinfo {author} {\bibfnamefont {D.~S.}\ \bibnamefont {Wang}},
  \bibinfo {author} {\bibfnamefont {C.~D.}\ \bibnamefont {Hill}}, \bibinfo
  {author} {\bibfnamefont {T.~D.}\ \bibnamefont {Ladd}}, \bibinfo {author}
  {\bibfnamefont {R.}~\bibnamefont {Van~Meter}}, \ and\ \bibinfo {author}
  {\bibfnamefont {L.~C.~L.}\ \bibnamefont {Hollenberg}},\ }\href {\doibase
  10.1103/PhysRevLett.104.180503} {\bibfield  {journal} {\bibinfo  {journal}
  {Phys. Rev. Lett.}\ }\textbf {\bibinfo {volume} {104}},\ \bibinfo {pages}
  {180503} (\bibinfo {year} {2010})}\BibitemShut {NoStop}%
\bibitem [{\citenamefont {Azuma}\ \emph {et~al.}(2015)\citenamefont {Azuma},
  \citenamefont {Tamaki},\ and\ \citenamefont {Lo}}]{azuma2015all}%
  \BibitemOpen
  \bibfield  {author} {\bibinfo {author} {\bibfnamefont {K.}~\bibnamefont
  {Azuma}}, \bibinfo {author} {\bibfnamefont {K.}~\bibnamefont {Tamaki}}, \
  and\ \bibinfo {author} {\bibfnamefont {H.-K.}\ \bibnamefont {Lo}},\ }\href
  {\doibase 10.1038/ncomms7787} {\bibfield  {journal} {\bibinfo  {journal}
  {Nature communications}\ }\textbf {\bibinfo {volume} {6}},\ \bibinfo {pages}
  {1} (\bibinfo {year} {2015})}\BibitemShut {NoStop}%
\bibitem [{\citenamefont {Lee}\ \emph {et~al.}(2019)\citenamefont {Lee},
  \citenamefont {Ralph},\ and\ \citenamefont {Jeong}}]{lee2019fundamental}%
  \BibitemOpen
  \bibfield  {author} {\bibinfo {author} {\bibfnamefont {S.-W.}\ \bibnamefont
  {Lee}}, \bibinfo {author} {\bibfnamefont {T.~C.}\ \bibnamefont {Ralph}}, \
  and\ \bibinfo {author} {\bibfnamefont {H.}~\bibnamefont {Jeong}},\ }\href
  {\doibase 10.1103/PhysRevA.100.052303} {\bibfield  {journal} {\bibinfo
  {journal} {Phys. Rev. A}\ }\textbf {\bibinfo {volume} {100}},\ \bibinfo
  {pages} {052303} (\bibinfo {year} {2019})}\BibitemShut {NoStop}%
\bibitem [{\citenamefont {Muralidharan}\ \emph {et~al.}(2014)\citenamefont
  {Muralidharan}, \citenamefont {Kim}, \citenamefont {L\"utkenhaus},
  \citenamefont {Lukin},\ and\ \citenamefont
  {Jiang}}]{muralidharan2014ultrafast}%
  \BibitemOpen
  \bibfield  {author} {\bibinfo {author} {\bibfnamefont {S.}~\bibnamefont
  {Muralidharan}}, \bibinfo {author} {\bibfnamefont {J.}~\bibnamefont {Kim}},
  \bibinfo {author} {\bibfnamefont {N.}~\bibnamefont {L\"utkenhaus}}, \bibinfo
  {author} {\bibfnamefont {M.~D.}\ \bibnamefont {Lukin}}, \ and\ \bibinfo
  {author} {\bibfnamefont {L.}~\bibnamefont {Jiang}},\ }\href {\doibase
  10.1103/PhysRevLett.112.250501} {\bibfield  {journal} {\bibinfo  {journal}
  {Phys. Rev. Lett.}\ }\textbf {\bibinfo {volume} {112}},\ \bibinfo {pages}
  {250501} (\bibinfo {year} {2014})}\BibitemShut {NoStop}%
\bibitem [{\citenamefont {Namiki}\ \emph {et~al.}(2016)\citenamefont {Namiki},
  \citenamefont {Jiang}, \citenamefont {Kim},\ and\ \citenamefont
  {L\"utkenhaus}}]{namiki2016role}%
  \BibitemOpen
  \bibfield  {author} {\bibinfo {author} {\bibfnamefont {R.}~\bibnamefont
  {Namiki}}, \bibinfo {author} {\bibfnamefont {L.}~\bibnamefont {Jiang}},
  \bibinfo {author} {\bibfnamefont {J.}~\bibnamefont {Kim}}, \ and\ \bibinfo
  {author} {\bibfnamefont {N.}~\bibnamefont {L\"utkenhaus}},\ }\href {\doibase
  10.1103/PhysRevA.94.052304} {\bibfield  {journal} {\bibinfo  {journal} {Phys.
  Rev. A}\ }\textbf {\bibinfo {volume} {94}},\ \bibinfo {pages} {052304}
  (\bibinfo {year} {2016})}\BibitemShut {NoStop}%
\bibitem [{\citenamefont {Muralidharan}\ \emph {et~al.}(2018)\citenamefont
  {Muralidharan}, \citenamefont {Zou}, \citenamefont {Li},\ and\ \citenamefont
  {Jiang}}]{muralidharan2018oneway}%
  \BibitemOpen
  \bibfield  {author} {\bibinfo {author} {\bibfnamefont {S.}~\bibnamefont
  {Muralidharan}}, \bibinfo {author} {\bibfnamefont {C.-L.}\ \bibnamefont
  {Zou}}, \bibinfo {author} {\bibfnamefont {L.}~\bibnamefont {Li}}, \ and\
  \bibinfo {author} {\bibfnamefont {L.}~\bibnamefont {Jiang}},\ }\href
  {\doibase 10.1103/PhysRevA.97.052316} {\bibfield  {journal} {\bibinfo
  {journal} {Phys. Rev. A}\ }\textbf {\bibinfo {volume} {97}},\ \bibinfo
  {pages} {052316} (\bibinfo {year} {2018})}\BibitemShut {NoStop}%
\bibitem [{\citenamefont {Ewert}\ \emph {et~al.}(2016)\citenamefont {Ewert},
  \citenamefont {Bergmann},\ and\ \citenamefont {van
  Loock}}]{ewert2016ultrafast}%
  \BibitemOpen
  \bibfield  {author} {\bibinfo {author} {\bibfnamefont {F.}~\bibnamefont
  {Ewert}}, \bibinfo {author} {\bibfnamefont {M.}~\bibnamefont {Bergmann}}, \
  and\ \bibinfo {author} {\bibfnamefont {P.}~\bibnamefont {van Loock}},\ }\href
  {\doibase 10.1103/PhysRevLett.117.210501} {\bibfield  {journal} {\bibinfo
  {journal} {Phys. Rev. Lett.}\ }\textbf {\bibinfo {volume} {117}},\ \bibinfo
  {pages} {210501} (\bibinfo {year} {2016})}\BibitemShut {NoStop}%
\bibitem [{\citenamefont {Ewert}\ and\ \citenamefont {van
  Loock}(2017)}]{ewert2017ultrafast}%
  \BibitemOpen
  \bibfield  {author} {\bibinfo {author} {\bibfnamefont {F.}~\bibnamefont
  {Ewert}}\ and\ \bibinfo {author} {\bibfnamefont {P.}~\bibnamefont {van
  Loock}},\ }\href {\doibase 10.1103/PhysRevA.95.012327} {\bibfield  {journal}
  {\bibinfo  {journal} {Phys. Rev. A}\ }\textbf {\bibinfo {volume} {95}},\
  \bibinfo {pages} {012327} (\bibinfo {year} {2017})}\BibitemShut {NoStop}%
\bibitem [{\citenamefont {Borregaard}\ \emph {et~al.}(2020)\citenamefont
  {Borregaard}, \citenamefont {Pichler}, \citenamefont {Schr{\"o}der},
  \citenamefont {Lukin}, \citenamefont {Lodahl},\ and\ \citenamefont
  {S{\o}rensen}}]{borregaard2020one}%
  \BibitemOpen
  \bibfield  {author} {\bibinfo {author} {\bibfnamefont {J.}~\bibnamefont
  {Borregaard}}, \bibinfo {author} {\bibfnamefont {H.}~\bibnamefont {Pichler}},
  \bibinfo {author} {\bibfnamefont {T.}~\bibnamefont {Schr{\"o}der}}, \bibinfo
  {author} {\bibfnamefont {M.~D.}\ \bibnamefont {Lukin}}, \bibinfo {author}
  {\bibfnamefont {P.}~\bibnamefont {Lodahl}}, \ and\ \bibinfo {author}
  {\bibfnamefont {A.~S.}\ \bibnamefont {S{\o}rensen}},\ }\href@noop {}
  {\bibfield  {journal} {\bibinfo  {journal} {Physical Review X}\ }\textbf
  {\bibinfo {volume} {10}},\ \bibinfo {pages} {021071} (\bibinfo {year}
  {2020})}\BibitemShut {NoStop}%
\bibitem [{\citenamefont {Zwerger}\ \emph {et~al.}(2016)\citenamefont
  {Zwerger}, \citenamefont {Briegel},\ and\ \citenamefont
  {D\"ur}}]{zwerger2016measurement}%
  \BibitemOpen
  \bibfield  {author} {\bibinfo {author} {\bibfnamefont {M.}~\bibnamefont
  {Zwerger}}, \bibinfo {author} {\bibfnamefont {H.~J.}\ \bibnamefont
  {Briegel}}, \ and\ \bibinfo {author} {\bibfnamefont {W.}~\bibnamefont
  {D\"ur}},\ }\href {\doibase 10.1007/s00340-015-6285-8} {\bibfield  {journal}
  {\bibinfo  {journal} {Appl. Phys. B}\ }\textbf {\bibinfo {volume} {122}},\
  \bibinfo {pages} {50} (\bibinfo {year} {2016})}\BibitemShut {NoStop}%
\bibitem [{\citenamefont {Panteleev}\ and\ \citenamefont
  {Kalachev}(2021{\natexlab{a}})}]{panteleev2021degenerate}%
  \BibitemOpen
  \bibfield  {author} {\bibinfo {author} {\bibfnamefont {P.}~\bibnamefont
  {Panteleev}}\ and\ \bibinfo {author} {\bibfnamefont {G.}~\bibnamefont
  {Kalachev}},\ }\href {\doibase 10.22331/q-2021-11-22-585} {\bibfield
  {journal} {\bibinfo  {journal} {{Quantum}}\ }\textbf {\bibinfo {volume}
  {5}},\ \bibinfo {pages} {585} (\bibinfo {year}
  {2021}{\natexlab{a}})}\BibitemShut {NoStop}%
\bibitem [{\citenamefont {Panteleev}\ and\ \citenamefont
  {Kalachev}(2021{\natexlab{b}})}]{panteleev2021quantum}%
  \BibitemOpen
  \bibfield  {author} {\bibinfo {author} {\bibfnamefont {P.}~\bibnamefont
  {Panteleev}}\ and\ \bibinfo {author} {\bibfnamefont {G.}~\bibnamefont
  {Kalachev}},\ }\href@noop {} {\bibfield  {journal} {\bibinfo  {journal} {IEEE
  Transactions on Information Theory}\ }\textbf {\bibinfo {volume} {68}},\
  \bibinfo {pages} {213} (\bibinfo {year} {2021}{\natexlab{b}})}\BibitemShut
  {NoStop}%
\bibitem [{\citenamefont {Varnava}\ \emph {et~al.}(2006)\citenamefont
  {Varnava}, \citenamefont {Browne},\ and\ \citenamefont
  {Rudolph}}]{varnava2006loss}%
  \BibitemOpen
  \bibfield  {author} {\bibinfo {author} {\bibfnamefont {M.}~\bibnamefont
  {Varnava}}, \bibinfo {author} {\bibfnamefont {D.~E.}\ \bibnamefont {Browne}},
  \ and\ \bibinfo {author} {\bibfnamefont {T.}~\bibnamefont {Rudolph}},\ }\href
  {\doibase 10.1103/PhysRevLett.97.120501} {\bibfield  {journal} {\bibinfo
  {journal} {Phys. Rev. Lett.}\ }\textbf {\bibinfo {volume} {97}},\ \bibinfo
  {pages} {120501} (\bibinfo {year} {2006})}\BibitemShut {NoStop}%
\bibitem [{\citenamefont {Raussendorf}\ \emph {et~al.}(2006)\citenamefont
  {Raussendorf}, \citenamefont {Harrington},\ and\ \citenamefont
  {Goyal}}]{raussendorf2006afault}%
  \BibitemOpen
  \bibfield  {author} {\bibinfo {author} {\bibfnamefont {R.}~\bibnamefont
  {Raussendorf}}, \bibinfo {author} {\bibfnamefont {J.}~\bibnamefont
  {Harrington}}, \ and\ \bibinfo {author} {\bibfnamefont {K.}~\bibnamefont
  {Goyal}},\ }\href {\doibase https://doi.org/10.1016/j.aop.2006.01.012}
  {\bibfield  {journal} {\bibinfo  {journal} {Annals of Physics}\ }\textbf
  {\bibinfo {volume} {321}},\ \bibinfo {pages} {2242} (\bibinfo {year}
  {2006})}\BibitemShut {NoStop}%
\bibitem [{\citenamefont {Bolt}\ \emph {et~al.}(2016)\citenamefont {Bolt},
  \citenamefont {Duclos-Cianci}, \citenamefont {Poulin},\ and\ \citenamefont
  {Stace}}]{bolt2016foliated}%
  \BibitemOpen
  \bibfield  {author} {\bibinfo {author} {\bibfnamefont {A.}~\bibnamefont
  {Bolt}}, \bibinfo {author} {\bibfnamefont {G.}~\bibnamefont {Duclos-Cianci}},
  \bibinfo {author} {\bibfnamefont {D.}~\bibnamefont {Poulin}}, \ and\ \bibinfo
  {author} {\bibfnamefont {T.~M.}\ \bibnamefont {Stace}},\ }\href {\doibase
  10.1103/PhysRevLett.117.070501} {\bibfield  {journal} {\bibinfo  {journal}
  {Phys. Rev. Lett.}\ }\textbf {\bibinfo {volume} {117}},\ \bibinfo {pages}
  {070501} (\bibinfo {year} {2016})}\BibitemShut {NoStop}%
\bibitem [{Note1()}]{Note1}%
  \BibitemOpen
  \bibinfo {note} {The intuitive reason behind this fact is that loss events
  can be corrected not only at the repeater node where the loss was observed
  but also through other stabilizers at neighboring repeater
  nodes.}\BibitemShut {Stop}%
\bibitem [{\citenamefont {Steane}(1996)}]{steane1996error}%
  \BibitemOpen
  \bibfield  {author} {\bibinfo {author} {\bibfnamefont {A.~M.}\ \bibnamefont
  {Steane}},\ }\href {\doibase 10.1103/PhysRevLett.77.793} {\bibfield
  {journal} {\bibinfo  {journal} {Phys. Rev. Lett.}\ }\textbf {\bibinfo
  {volume} {77}},\ \bibinfo {pages} {793} (\bibinfo {year} {1996})}\BibitemShut
  {NoStop}%
\bibitem [{\citenamefont {Brendel}\ \emph {et~al.}(1999)\citenamefont
  {Brendel}, \citenamefont {Gisin}, \citenamefont {Tittel},\ and\ \citenamefont
  {Zbinden}}]{brendel1999pulsed}%
  \BibitemOpen
  \bibfield  {author} {\bibinfo {author} {\bibfnamefont {J.}~\bibnamefont
  {Brendel}}, \bibinfo {author} {\bibfnamefont {N.}~\bibnamefont {Gisin}},
  \bibinfo {author} {\bibfnamefont {W.}~\bibnamefont {Tittel}}, \ and\ \bibinfo
  {author} {\bibfnamefont {H.}~\bibnamefont {Zbinden}},\ }\href {\doibase
  10.1103/PhysRevLett.82.2594} {\bibfield  {journal} {\bibinfo  {journal}
  {Phys. Rev. Lett.}\ }\textbf {\bibinfo {volume} {82}},\ \bibinfo {pages}
  {2594} (\bibinfo {year} {1999})}\BibitemShut {NoStop}%
\bibitem [{\citenamefont {Knill}\ \emph {et~al.}(2001)\citenamefont {Knill},
  \citenamefont {Laflamme},\ and\ \citenamefont {Milburn}}]{knill2001scheme}%
  \BibitemOpen
  \bibfield  {author} {\bibinfo {author} {\bibfnamefont {E.}~\bibnamefont
  {Knill}}, \bibinfo {author} {\bibfnamefont {R.}~\bibnamefont {Laflamme}}, \
  and\ \bibinfo {author} {\bibfnamefont {G.~J.}\ \bibnamefont {Milburn}},\
  }\href@noop {} {\bibfield  {journal} {\bibinfo  {journal} {nature}\ }\textbf
  {\bibinfo {volume} {409}},\ \bibinfo {pages} {46} (\bibinfo {year}
  {2001})}\BibitemShut {NoStop}%
\bibitem [{\citenamefont {Knill}(2005)}]{knill2005quantum}%
  \BibitemOpen
  \bibfield  {author} {\bibinfo {author} {\bibfnamefont {E.}~\bibnamefont
  {Knill}},\ }\href {\doibase 10.1038/nature03350} {\bibfield  {journal}
  {\bibinfo  {journal} {Nature}\ }\textbf {\bibinfo {volume} {434}},\ \bibinfo
  {pages} {39} (\bibinfo {year} {2005})}\BibitemShut {NoStop}%
\bibitem [{\citenamefont {Hein}\ \emph {et~al.}(2006)\citenamefont {Hein},
  \citenamefont {D\"ur}, \citenamefont {Eisert}, \citenamefont {Raussendorf},
  \citenamefont {den Nest},\ and\ \citenamefont
  {Briegel}}]{hein2006entanglement}%
  \BibitemOpen
  \bibfield  {author} {\bibinfo {author} {\bibfnamefont {M.}~\bibnamefont
  {Hein}}, \bibinfo {author} {\bibfnamefont {W.}~\bibnamefont {D\"ur}},
  \bibinfo {author} {\bibfnamefont {J.}~\bibnamefont {Eisert}}, \bibinfo
  {author} {\bibfnamefont {R.}~\bibnamefont {Raussendorf}}, \bibinfo {author}
  {\bibfnamefont {M.~V.}\ \bibnamefont {den Nest}}, \ and\ \bibinfo {author}
  {\bibfnamefont {H.-J.}\ \bibnamefont {Briegel}},\ }\href@noop {} {\bibfield
  {journal} {\bibinfo  {journal} {arXiv:quant-ph/0602096}\ } (\bibinfo {year}
  {2006})}\BibitemShut {NoStop}%
\bibitem [{\citenamefont {Raussendorf}\ \emph {et~al.}(2003)\citenamefont
  {Raussendorf}, \citenamefont {Browne},\ and\ \citenamefont
  {Briegel}}]{raussendorf2003measurement}%
  \BibitemOpen
  \bibfield  {author} {\bibinfo {author} {\bibfnamefont {R.}~\bibnamefont
  {Raussendorf}}, \bibinfo {author} {\bibfnamefont {D.~E.}\ \bibnamefont
  {Browne}}, \ and\ \bibinfo {author} {\bibfnamefont {H.~J.}\ \bibnamefont
  {Briegel}},\ }\href {\doibase 10.1103/PhysRevA.68.022312} {\bibfield
  {journal} {\bibinfo  {journal} {Phys. Rev. A}\ }\textbf {\bibinfo {volume}
  {68}},\ \bibinfo {pages} {022312} (\bibinfo {year} {2003})}\BibitemShut
  {NoStop}%
\bibitem [{\citenamefont {Rozp{\k{e}}dek}\ \emph {et~al.}(2021)\citenamefont
  {Rozp{\k{e}}dek}, \citenamefont {Noh}, \citenamefont {Xu}, \citenamefont
  {Guha},\ and\ \citenamefont {Jiang}}]{rozpedek2021quantum}%
  \BibitemOpen
  \bibfield  {author} {\bibinfo {author} {\bibfnamefont {F.}~\bibnamefont
  {Rozp{\k{e}}dek}}, \bibinfo {author} {\bibfnamefont {K.}~\bibnamefont {Noh}},
  \bibinfo {author} {\bibfnamefont {Q.}~\bibnamefont {Xu}}, \bibinfo {author}
  {\bibfnamefont {S.}~\bibnamefont {Guha}}, \ and\ \bibinfo {author}
  {\bibfnamefont {L.}~\bibnamefont {Jiang}},\ }\href {\doibase
  10.1038/s41534-021-00438-7} {\bibfield  {journal} {\bibinfo  {journal} {npj
  Quantum Information}\ }\textbf {\bibinfo {volume} {7}} (\bibinfo {year}
  {2021}),\ 10.1038/s41534-021-00438-7}\BibitemShut {NoStop}%
\bibitem [{\citenamefont {Lindner}\ and\ \citenamefont
  {Rudolph}(2009)}]{lindner2009proposal}%
  \BibitemOpen
  \bibfield  {author} {\bibinfo {author} {\bibfnamefont {N.~H.}\ \bibnamefont
  {Lindner}}\ and\ \bibinfo {author} {\bibfnamefont {T.}~\bibnamefont
  {Rudolph}},\ }\href {\doibase 10.1103/PhysRevLett.103.113602} {\bibfield
  {journal} {\bibinfo  {journal} {Phys. Rev. Lett.}\ }\textbf {\bibinfo
  {volume} {103}},\ \bibinfo {pages} {113602} (\bibinfo {year}
  {2009})}\BibitemShut {NoStop}%
\bibitem [{\citenamefont {Buterakos}\ \emph {et~al.}(2017)\citenamefont
  {Buterakos}, \citenamefont {Barnes},\ and\ \citenamefont
  {Economou}}]{buterakos2017deterministic}%
  \BibitemOpen
  \bibfield  {author} {\bibinfo {author} {\bibfnamefont {D.}~\bibnamefont
  {Buterakos}}, \bibinfo {author} {\bibfnamefont {E.}~\bibnamefont {Barnes}}, \
  and\ \bibinfo {author} {\bibfnamefont {S.~E.}\ \bibnamefont {Economou}},\
  }\href {\doibase 10.1103/PhysRevX.7.041023} {\bibfield  {journal} {\bibinfo
  {journal} {Phys. Rev. X}\ }\textbf {\bibinfo {volume} {7}},\ \bibinfo {pages}
  {041023} (\bibinfo {year} {2017})}\BibitemShut {NoStop}%
\bibitem [{\citenamefont {Pichler}\ and\ \citenamefont
  {Zoller}(2016)}]{pichler2016photonic}%
  \BibitemOpen
  \bibfield  {author} {\bibinfo {author} {\bibfnamefont {H.}~\bibnamefont
  {Pichler}}\ and\ \bibinfo {author} {\bibfnamefont {P.}~\bibnamefont
  {Zoller}},\ }\href {\doibase 10.1103/PhysRevLett.116.093601} {\bibfield
  {journal} {\bibinfo  {journal} {Phys. Rev. Lett.}\ }\textbf {\bibinfo
  {volume} {116}},\ \bibinfo {pages} {093601} (\bibinfo {year}
  {2016})}\BibitemShut {NoStop}%
\bibitem [{\citenamefont {Pichler}\ \emph {et~al.}(2017)\citenamefont
  {Pichler}, \citenamefont {Choi}, \citenamefont {Zoller},\ and\ \citenamefont
  {Lukin}}]{pichler2017universal}%
  \BibitemOpen
  \bibfield  {author} {\bibinfo {author} {\bibfnamefont {H.}~\bibnamefont
  {Pichler}}, \bibinfo {author} {\bibfnamefont {S.}~\bibnamefont {Choi}},
  \bibinfo {author} {\bibfnamefont {P.}~\bibnamefont {Zoller}}, \ and\ \bibinfo
  {author} {\bibfnamefont {M.~D.}\ \bibnamefont {Lukin}},\ }\href {\doibase
  10.1073/pnas.1711003114} {\bibfield  {journal} {\bibinfo  {journal}
  {Proceedings of the National Academy of Sciences}\ }\textbf {\bibinfo
  {volume} {114}},\ \bibinfo {pages} {11362} (\bibinfo {year}
  {2017})}\BibitemShut {NoStop}%
\bibitem [{\citenamefont {Wan}\ \emph {et~al.}(2021)\citenamefont {Wan},
  \citenamefont {Choi}, \citenamefont {Kim}, \citenamefont {Shutty},\ and\
  \citenamefont {Hayden}}]{wan2021fault}%
  \BibitemOpen
  \bibfield  {author} {\bibinfo {author} {\bibfnamefont {K.}~\bibnamefont
  {Wan}}, \bibinfo {author} {\bibfnamefont {S.}~\bibnamefont {Choi}}, \bibinfo
  {author} {\bibfnamefont {I.~H.}\ \bibnamefont {Kim}}, \bibinfo {author}
  {\bibfnamefont {N.}~\bibnamefont {Shutty}}, \ and\ \bibinfo {author}
  {\bibfnamefont {P.}~\bibnamefont {Hayden}},\ }\href {\doibase
  10.1103/PRXQuantum.2.040345} {\bibfield  {journal} {\bibinfo  {journal} {PRX
  Quantum}\ }\textbf {\bibinfo {volume} {2}},\ \bibinfo {pages} {040345}
  (\bibinfo {year} {2021})}\BibitemShut {NoStop}%
\bibitem [{\citenamefont {Zhan}\ and\ \citenamefont
  {Sun}(2020)}]{zhan2020deterministic}%
  \BibitemOpen
  \bibfield  {author} {\bibinfo {author} {\bibfnamefont {Y.}~\bibnamefont
  {Zhan}}\ and\ \bibinfo {author} {\bibfnamefont {S.}~\bibnamefont {Sun}},\
  }\href {\doibase 10.1103/PhysRevLett.125.223601} {\bibfield  {journal}
  {\bibinfo  {journal} {Phys. Rev. Lett.}\ }\textbf {\bibinfo {volume} {125}},\
  \bibinfo {pages} {223601} (\bibinfo {year} {2020})}\BibitemShut {NoStop}%
\bibitem [{\citenamefont {Shapourian}\ and\ \citenamefont
  {Shabani}(2022)}]{shapourian2022modular}%
  \BibitemOpen
  \bibfield  {author} {\bibinfo {author} {\bibfnamefont {H.}~\bibnamefont
  {Shapourian}}\ and\ \bibinfo {author} {\bibfnamefont {A.}~\bibnamefont
  {Shabani}},\ }\href {\doibase https://arxiv.org/abs/2206.11307} {\bibfield
  {journal} {\bibinfo  {journal} {arXiv:2206.11307}\ } (\bibinfo {year}
  {2022}),\ https://arxiv.org/abs/2206.11307}\BibitemShut {NoStop}%
\bibitem [{\citenamefont {Li}\ \emph {et~al.}(2022)\citenamefont {Li},
  \citenamefont {Economou},\ and\ \citenamefont {Barnes}}]{li2022photonic}%
  \BibitemOpen
  \bibfield  {author} {\bibinfo {author} {\bibfnamefont {B.}~\bibnamefont
  {Li}}, \bibinfo {author} {\bibfnamefont {S.~E.}\ \bibnamefont {Economou}}, \
  and\ \bibinfo {author} {\bibfnamefont {E.}~\bibnamefont {Barnes}},\ }\href
  {\doibase 10.1038/s41534-022-00562-y} {\bibfield  {journal} {\bibinfo
  {journal} {npj Quantum Information}\ }\textbf {\bibinfo {volume} {8}},\
  \bibinfo {pages} {1} (\bibinfo {year} {2022})}\BibitemShut {NoStop}%
\bibitem [{\citenamefont {Schrinski}\ \emph {et~al.}(2021)\citenamefont
  {Schrinski}, \citenamefont {Lamaison},\ and\ \citenamefont
  {S\o{}rensen}}]{schrinski2021passive}%
  \BibitemOpen
  \bibfield  {author} {\bibinfo {author} {\bibfnamefont {B.}~\bibnamefont
  {Schrinski}}, \bibinfo {author} {\bibfnamefont {M.}~\bibnamefont {Lamaison}},
  \ and\ \bibinfo {author} {\bibfnamefont {A.~S.}\ \bibnamefont
  {S\o{}rensen}},\ }\href {https://arxiv.org/abs/2112.11328} {\bibfield
  {journal} {\bibinfo  {journal} {arXiv:2112.11328}\ } (\bibinfo {year}
  {2021})}\BibitemShut {NoStop}%
\bibitem [{\citenamefont {Browne}\ and\ \citenamefont
  {Rudolph}(2005)}]{browne2005resource}%
  \BibitemOpen
  \bibfield  {author} {\bibinfo {author} {\bibfnamefont {D.~E.}\ \bibnamefont
  {Browne}}\ and\ \bibinfo {author} {\bibfnamefont {T.}~\bibnamefont
  {Rudolph}},\ }\href {\doibase 10.1103/PhysRevLett.95.010501} {\bibfield
  {journal} {\bibinfo  {journal} {Phys. Rev. Lett.}\ }\textbf {\bibinfo
  {volume} {95}},\ \bibinfo {pages} {010501} (\bibinfo {year}
  {2005})}\BibitemShut {NoStop}%
\bibitem [{\citenamefont {Bartolucci}\ \emph {et~al.}(2021)\citenamefont
  {Bartolucci}, \citenamefont {Birchall}, \citenamefont {Bombin}, \citenamefont
  {Cable}, \citenamefont {Dawson}, \citenamefont {Gimeno-Segovia},
  \citenamefont {Johnston}, \citenamefont {Kieling}, \citenamefont {Nickerson},
  \citenamefont {Pant} \emph {et~al.}}]{bartolucci2021fusion}%
  \BibitemOpen
  \bibfield  {author} {\bibinfo {author} {\bibfnamefont {S.}~\bibnamefont
  {Bartolucci}}, \bibinfo {author} {\bibfnamefont {P.}~\bibnamefont
  {Birchall}}, \bibinfo {author} {\bibfnamefont {H.}~\bibnamefont {Bombin}},
  \bibinfo {author} {\bibfnamefont {H.}~\bibnamefont {Cable}}, \bibinfo
  {author} {\bibfnamefont {C.}~\bibnamefont {Dawson}}, \bibinfo {author}
  {\bibfnamefont {M.}~\bibnamefont {Gimeno-Segovia}}, \bibinfo {author}
  {\bibfnamefont {E.}~\bibnamefont {Johnston}}, \bibinfo {author}
  {\bibfnamefont {K.}~\bibnamefont {Kieling}}, \bibinfo {author} {\bibfnamefont
  {N.}~\bibnamefont {Nickerson}}, \bibinfo {author} {\bibfnamefont
  {M.}~\bibnamefont {Pant}},  \emph {et~al.},\ }\href
  {https://arxiv.org/abs/2101.09310} {\bibfield  {journal} {\bibinfo  {journal}
  {arXiv:2101.09310}\ } (\bibinfo {year} {2021})}\BibitemShut {NoStop}%
\bibitem [{\citenamefont {Wang}\ \emph {et~al.}(2020)\citenamefont {Wang},
  \citenamefont {Sciarrino}, \citenamefont {Laing},\ and\ \citenamefont
  {Thompson}}]{wang2020integrated}%
  \BibitemOpen
  \bibfield  {author} {\bibinfo {author} {\bibfnamefont {J.}~\bibnamefont
  {Wang}}, \bibinfo {author} {\bibfnamefont {F.}~\bibnamefont {Sciarrino}},
  \bibinfo {author} {\bibfnamefont {A.}~\bibnamefont {Laing}}, \ and\ \bibinfo
  {author} {\bibfnamefont {M.~G.}\ \bibnamefont {Thompson}},\ }\href {\doibase
  10.1038/s41566-019-0532-1} {\bibfield  {journal} {\bibinfo  {journal} {Nature
  Photonics}\ }\textbf {\bibinfo {volume} {14}},\ \bibinfo {pages} {273}
  (\bibinfo {year} {2020})}\BibitemShut {NoStop}%
\bibitem [{\citenamefont {Ewert}\ and\ \citenamefont {van
  Loock}(2014)}]{ewert2014efficient}%
  \BibitemOpen
  \bibfield  {author} {\bibinfo {author} {\bibfnamefont {F.}~\bibnamefont
  {Ewert}}\ and\ \bibinfo {author} {\bibfnamefont {P.}~\bibnamefont {van
  Loock}},\ }\href {\doibase 10.1103/PhysRevLett.113.140403} {\bibfield
  {journal} {\bibinfo  {journal} {Phys. Rev. Lett.}\ }\textbf {\bibinfo
  {volume} {113}},\ \bibinfo {pages} {140403} (\bibinfo {year}
  {2014})}\BibitemShut {NoStop}%
\bibitem [{\citenamefont {Bhaskar}\ \emph {et~al.}(2020)\citenamefont
  {Bhaskar}, \citenamefont {Riedinger}, \citenamefont {Machielse},
  \citenamefont {Levonian}, \citenamefont {Nguyen}, \citenamefont {Knall},
  \citenamefont {Park}, \citenamefont {Englund}, \citenamefont {Lon{\v{c}}ar},
  \citenamefont {Sukachev} \emph {et~al.}}]{bhaskar2020experimental}%
  \BibitemOpen
  \bibfield  {author} {\bibinfo {author} {\bibfnamefont {M.~K.}\ \bibnamefont
  {Bhaskar}}, \bibinfo {author} {\bibfnamefont {R.}~\bibnamefont {Riedinger}},
  \bibinfo {author} {\bibfnamefont {B.}~\bibnamefont {Machielse}}, \bibinfo
  {author} {\bibfnamefont {D.~S.}\ \bibnamefont {Levonian}}, \bibinfo {author}
  {\bibfnamefont {C.~T.}\ \bibnamefont {Nguyen}}, \bibinfo {author}
  {\bibfnamefont {E.~N.}\ \bibnamefont {Knall}}, \bibinfo {author}
  {\bibfnamefont {H.}~\bibnamefont {Park}}, \bibinfo {author} {\bibfnamefont
  {D.}~\bibnamefont {Englund}}, \bibinfo {author} {\bibfnamefont
  {M.}~\bibnamefont {Lon{\v{c}}ar}}, \bibinfo {author} {\bibfnamefont {D.~D.}\
  \bibnamefont {Sukachev}},  \emph {et~al.},\ }\href@noop {} {\bibfield
  {journal} {\bibinfo  {journal} {Nature}\ }\textbf {\bibinfo {volume} {580}},\
  \bibinfo {pages} {60} (\bibinfo {year} {2020})}\BibitemShut {NoStop}%
\bibitem [{\citenamefont {Hilaire}\ \emph {et~al.}(2022)\citenamefont
  {Hilaire}, \citenamefont {Vidro}, \citenamefont {Eisenberg},\ and\
  \citenamefont {Economou}}]{hilaire2022near}%
  \BibitemOpen
  \bibfield  {author} {\bibinfo {author} {\bibfnamefont {P.}~\bibnamefont
  {Hilaire}}, \bibinfo {author} {\bibfnamefont {L.}~\bibnamefont {Vidro}},
  \bibinfo {author} {\bibfnamefont {H.~S.}\ \bibnamefont {Eisenberg}}, \ and\
  \bibinfo {author} {\bibfnamefont {S.~E.}\ \bibnamefont {Economou}},\ }\href
  {https://arxiv.org/abs/2205.09750} {\bibfield  {journal} {\bibinfo  {journal}
  {arXiv:2205.09750}\ } (\bibinfo {year} {2022})}\BibitemShut {NoStop}%
\bibitem [{\citenamefont {Delfosse}\ and\ \citenamefont
  {Z\'emor}(2020)}]{delfosse2020linear}%
  \BibitemOpen
  \bibfield  {author} {\bibinfo {author} {\bibfnamefont {N.}~\bibnamefont
  {Delfosse}}\ and\ \bibinfo {author} {\bibfnamefont {G.}~\bibnamefont
  {Z\'emor}},\ }\href {\doibase 10.1103/PhysRevResearch.2.033042} {\bibfield
  {journal} {\bibinfo  {journal} {Phys. Rev. Research}\ }\textbf {\bibinfo
  {volume} {2}},\ \bibinfo {pages} {033042} (\bibinfo {year}
  {2020})}\BibitemShut {NoStop}%
\bibitem [{\citenamefont {Connolly}\ \emph {et~al.}(2022)\citenamefont
  {Connolly}, \citenamefont {Londe}, \citenamefont {Leverrier},\ and\
  \citenamefont {Delfosse}}]{connolly2022fast}%
  \BibitemOpen
  \bibfield  {author} {\bibinfo {author} {\bibfnamefont {N.}~\bibnamefont
  {Connolly}}, \bibinfo {author} {\bibfnamefont {V.}~\bibnamefont {Londe}},
  \bibinfo {author} {\bibfnamefont {A.}~\bibnamefont {Leverrier}}, \ and\
  \bibinfo {author} {\bibfnamefont {N.}~\bibnamefont {Delfosse}},\ }\href
  {https://arxiv.org/abs/2208.01002} {\bibfield  {journal} {\bibinfo  {journal}
  {arXiv:2208.01002}\ } (\bibinfo {year} {2022})}\BibitemShut {NoStop}%
\bibitem [{\citenamefont {Fukui}\ \emph {et~al.}(2021)\citenamefont {Fukui},
  \citenamefont {Alexander},\ and\ \citenamefont {van Loock}}]{fukui2021all}%
  \BibitemOpen
  \bibfield  {author} {\bibinfo {author} {\bibfnamefont {K.}~\bibnamefont
  {Fukui}}, \bibinfo {author} {\bibfnamefont {R.~N.}\ \bibnamefont
  {Alexander}}, \ and\ \bibinfo {author} {\bibfnamefont {P.}~\bibnamefont {van
  Loock}},\ }\href {\doibase 10.1103/PhysRevResearch.3.033118} {\bibfield
  {journal} {\bibinfo  {journal} {Phys. Rev. Research}\ }\textbf {\bibinfo
  {volume} {3}},\ \bibinfo {pages} {033118} (\bibinfo {year}
  {2021})}\BibitemShut {NoStop}%
\bibitem [{\citenamefont {Raveendran}\ \emph {et~al.}(2022)\citenamefont
  {Raveendran}, \citenamefont {Rengaswamy}, \citenamefont {Rozp{\k{e}}dek},
  \citenamefont {Raina}, \citenamefont {Jiang},\ and\ \citenamefont
  {Vasi{\'c}}}]{raveendran2022finite}%
  \BibitemOpen
  \bibfield  {author} {\bibinfo {author} {\bibfnamefont {N.}~\bibnamefont
  {Raveendran}}, \bibinfo {author} {\bibfnamefont {N.}~\bibnamefont
  {Rengaswamy}}, \bibinfo {author} {\bibfnamefont {F.}~\bibnamefont
  {Rozp{\k{e}}dek}}, \bibinfo {author} {\bibfnamefont {A.}~\bibnamefont
  {Raina}}, \bibinfo {author} {\bibfnamefont {L.}~\bibnamefont {Jiang}}, \ and\
  \bibinfo {author} {\bibfnamefont {B.}~\bibnamefont {Vasi{\'c}}},\ }\href@noop
  {} {\bibfield  {journal} {\bibinfo  {journal} {Quantum}\ }\textbf {\bibinfo
  {volume} {6}},\ \bibinfo {pages} {767} (\bibinfo {year} {2022})}\BibitemShut
  {NoStop}%
\bibitem [{Note5()}]{Note5}%
  \BibitemOpen
  \bibinfo {note} {While such protocols feature no need for a quantum memory,
  their realization is largely hindered by the difficulty of generating GKP
  states in the optical domain with large squeezing values.}\BibitemShut
  {Stop}%
\bibitem [{\citenamefont {Gottesman}(1997)}]{gottesman1997stabilizer}%
  \BibitemOpen
  \bibfield  {author} {\bibinfo {author} {\bibfnamefont {D.}~\bibnamefont
  {Gottesman}},\ }\href@noop {} {\emph {\bibinfo {title} {Stabilizer codes and
  quantum error correction}}}\ (\bibinfo  {publisher} {California Institute of
  Technology},\ \bibinfo {year} {1997})\BibitemShut {NoStop}%
\bibitem [{\citenamefont {Nielsen}\ and\ \citenamefont
  {Chuang}(2010)}]{Nielsen-Chuang}%
  \BibitemOpen
  \bibfield  {author} {\bibinfo {author} {\bibfnamefont {M.~A.}\ \bibnamefont
  {Nielsen}}\ and\ \bibinfo {author} {\bibfnamefont {I.~L.}\ \bibnamefont
  {Chuang}},\ }\href {\doibase 10.1017/CBO9780511976667} {\emph {\bibinfo
  {title} {Quantum Computation and Quantum Information: 10th Anniversary
  Edition}}}\ (\bibinfo  {publisher} {Cambridge Univ. Pr., Cambridge, UK},\
  \bibinfo {year} {2010})\BibitemShut {NoStop}%
\bibitem [{\citenamefont {Kovalev}\ and\ \citenamefont
  {Pryadko}(2013)}]{Kovalev2013QLDPCfiniterate}%
  \BibitemOpen
  \bibfield  {author} {\bibinfo {author} {\bibfnamefont {A.~A.}\ \bibnamefont
  {Kovalev}}\ and\ \bibinfo {author} {\bibfnamefont {L.~P.}\ \bibnamefont
  {Pryadko}},\ }\href {\doibase 10.1103/PhysRevA.88.012311} {\bibfield
  {journal} {\bibinfo  {journal} {Phys. Rev. A}\ }\textbf {\bibinfo {volume}
  {88}},\ \bibinfo {pages} {012311} (\bibinfo {year} {2013})}\BibitemShut
  {NoStop}%
\end{thebibliography}%

\end{document}